\documentclass[12pt,preprint,tighten]{aastex}
\usepackage[]{times, graphicx}
\usepackage{epsfig}                     
\usepackage{graphicx}                    

\newcommand{\pref}{\protect\ref}
\newcommand{\trace}{TRACE}
\newcommand{\sdo}{SDO}

\slugcomment{In Press Solar Physics}
\begin{document}
\title{A Coherence-Based Approach for Tracking Waves in the Solar Corona}
\author{Scott W.~McIntosh$^{1,3}$, Bart De Pontieu$^2$ \& Steven Tomczyk$^1$}
\affil{1: High Altitude Observatory, National Center for Atmospheric Research,\\P.O. Box 3000, Boulder, CO 80307, USA.}
\affil{2: Lockheed Martin Solar and Astrophysics Lab,\\ 3251 Hanover St., Org. ADBS, Bldg. 252, Palo Alto, CA  94304, USA.}
\affil{3: {\url mailto:mscott@ucar.edu}}
%
\shortauthors{S.W., McIntosh {\it et~al.}}
\shorttitle{Automated Coronal Wave Tracking}

%


\begin{abstract}
We consider the problem of automatically (and robustly) isolating and extracting information about waves and oscillations observed in EUV image sequences of the solar corona with a view to near real-time application to data from the Atmospheric Imaging Array (AIA) on the {\em Solar Dynamics Observatory} (\sdo). We find that a simple coherence / travel-time based approach detects and provides a wealth of information on transverse and longitudinal wave phenomena in the test sequences provided by the {\em Transition Region and Coronal Explorer} (\trace). The results of the search are ÒprunedÓ (based on diagnostic errors) to minimize false-detections such that the remainder provides robust measurements of waves in the solar corona, with the calculated propagation speed allowing automated distinction between various wave modes. In this paper we discuss the technique, present results on the \trace{} test sequences, and describe how our method can be used to automatically process the enormous flow of data ($\approx$1Tb/day) that will be provided by \sdo/AIA after launch in late 2008.
\end{abstract}

%


%

Since the launch of the {\em Transition Region and Coronal Explorer} \citep[\trace;][]{1999SoPh..187..229H} the community has invested a great deal of effort on the identification and analysis of longitudinal and transverse wave phenomena in loop structures seen in its EUV images of the corona \citep[see][for a good overview]{2005LRSP....2....3N}. The great interest in finding and characterizing coronal waves is driven by the promise of coronal seismology which could lead to the determination of otherwise inaccessible physical properties of the solar atmosphere by studying phase speeds, amplitudes, dissipation, {\it etc.} of the observed waves \citep[][]{2005LRSP....2....3N}.

There are many unresolved issues in coronal seismology: for example, it is unclear why only a subset of coronal loops show transverse oscillations in the wake of a flare or CME \citep[{\it e.g.},][]{2002SoPh..206...99A}, which physical processes dominate the damping of these transverse oscillations \citep[{\it e.g.},][]{2002ApJ...576L.153O, 2007AAP...463..333A, 2007SoPh..246..231T}, why only a subset of coronal loops show propagating slow-mode oscillations \citep[{\it e.g.},][]{2000AAP...355L..23D, 2007SoPh..246...53D}, how the propagation of slow-mode waves at different temperatures can be used to probe the thermal structure and loop scale heights \citep[][]{2003AAP...404L...1K}, or how lower-atmospheric oscillations generally leak into the corona \citep[{\it e.g.},][]{2005ApJ...624L..61D, 2006ApJ...648L.151J, 2008ApJ...676L..85K}?

Many of these issues require more extensive study with much larger statistical samples. It is surprising that the number of wave and oscillation detections in the above papers is only of order several dozen for both longitudinal and transverse oscillations. This is a very small number compared to the volume of \trace{} data given that flares occur often and ``wave-leakage'' from the solar interior to the lower portions of the outer solar atmosphere is apparently quite abundant \citep[][]{2005ApJ...624L..61D,2006ApJ...648L.151J,2008ApJ...676L..85K}. The discrepancy between the plethora of observed wave driven phenomena in the chromosphere and transition region, and the relative paucity of waves seen in the corona partly motivates the effort discussed herein. Is the small number of coronal wave observations caused by the fact that all of the \trace{} observations of waves have been found as a result of a ``manual'' search, {\it i.e.}, limited by human patience and detection skills? Or are there actually very few locations in the corona where waves can be observed? 

One of our aims is to develop an automated wave-detection algorithm to help expand the number of known oscillations, and build up significant statistics on how frequent coronal oscillations occur. Such larger statistical samples are crucial to address many of the unresolved issues in coronal seismology, and will be within reach with the Atmospheric Imaging Assembly (AIA) after launch of the {\em Solar Dynamics Observatory} (\sdo) by 2009. This instrument will provide full-disk coronal imaging at significantly increased signal-to-noise in eight coronal wavelengths at ten-second cadence continuously, producing an enormous data flow of $\approx$1Tb day$^{-1}$. The potential for significantly increased detection of coronal oscillations and waves with such a data rate and quality are very high. However, AIA's enormous data rate renders business-as-usual approaches that involve manually looking through data sets and individually flagging events unfeasible. Since it is unpractical to transfer the full volume of AIA data to remote users, it is critical that automated algorithms be developed that can search for wave-like phenomena and automatically flag temporal and spatial regions of interest for later, more detailed analysis. There have been a few papers recently  that have started addressing some of the issues that need to be resolved to enable full optimal use of the AIA data, such as automated detection of locations with significant oscillatory power \citep[][]{2007SoPh..241..397N}, or semi-automated detection of propagating and standing waves \citep[][]{2007SoPh..tmp..102S}. The approach presented in this paper may be the first that can reliably and automatically distinguish between longitudinal and transverse oscillations at the same time as rejecting most false positive detections of oscillations. Our approach is based on the technique used to analyze and detect coronal Alfv\'{e}n waves with the CoMP instrument \citep[][]{2007Sci...317.1192T}.

In the following section we discuss the method developed, illustrating its application on example \trace{} datasets (see {\it Sect.}~\pref{s:data}) which are known to show a host of transverse and longitudinal wave phenomena and are therefore good for testing our approach. In {\it Sect.}~\pref{s:results} we explore the results of the analysis on the test data and discuss the method employed to minimize false detections. In {\it Sect.}~\pref{s:discuss} we compare some of our results with previous analyses of the \trace{} datasets we have used, and discuss how our method could be used in a practical manner for \sdo/AIA data.

\section{Method}\label{s:method}
We adapt the phase travel-time analysis \citep[{\it e.g.},][]{1994ApJ...434..795J,1997ApJ...485L..49J, 2004ApJ...613L.185F,2004ApJ...609L..95M} to isolate the propagation characteristics of the wave modes observed in the \trace{} image sequences time-series as a function of frequency. This technique is performed as a Fourier comparison of time-series in neighboring image pixels to extract information about the cross-correlation, coherence, cross-power and phase dependence of the neighboring signals. As demonstrated in \citet{2007Sci...317.1192T} this is a particularly simple, but powerful technique to track coronal waves, albeit in the plane-of-the-sky. The technique has the inherent flexibility to be employed rapidly in the spectral domain producing robust results.

The first step in the travel-time analysis involves computing the multi dimensional Fourier transform of the EUV image cube ($I(x,y,t)$) in the third dimension which gives a complex cube $I^{\ast}_{f}(x, y, f)$ that now has frequency $f$ [units: mHz] as its third dimension. Before calculating the Fourier transform we apply a Hanning window on the timeseries to reduce the impact of temporal trends in the short data sequences.

The analysis involves comparisons of the intensity timeseries of our pixel of interest $(x,y)$ and that of a pixel within a square region (of dimensions $dx \times dy$)  around the central pixel. At this point the analysis diverges into the spectral ({\it Sect.}~\pref{s:sda}) and temporal domains ({\it Sect.}~\pref{s:ptr}) where the former has been developed by \citet{Tomczyk2008} to supersede the analysis of \citet{2007Sci...317.1192T}. We have chosen the range to be $\pm$10 pixels in both spatial dimensions.

\subsection{Spectral Domain Analysis}\label{s:sda}

For the range of pixels $[x \pm dx : y \pm dy]$ we compute the cross-spectrum
\begin{equation}
CS_{RB}(f) = I^{\ast}_{R}(f) * I^{\ast}_{B}(f)^{\dagger}
\end{equation}
 between the Fourier transformed timeseries at the reference pixel $I^{\ast}_{R}(f)$ for any other pixel in the analysis box $I^{\ast}_{B}(f)$, with $I(f)^{\dagger}$ being the complex conjugate of $I(f)$. We then smooth the cross-spectrum with a five-point boxcar smoothing function in frequency to reduce the contribution of noise in the computation. In the spectral domain we wish to rapidly compute wave diagnostics in several characteristic frequency filters simultaneously and for the sake of simplicity we will use Gaussian filters of the form $G(f_{0}, \delta f)$ where $f_{0}$ is the central frequency and $\delta f$ is the width of the Gaussian filter chosen which, for this paper, is $f_{0}$/10. So, for each filter selected we can simultaneously compute the weighted signal cross-power [$WCP_{RB}(f_{0})$], phase [$\phi_{RB}(f_{0})$] , coherence [$C_{RB}(f_{0})$] and phase travel-time [$T_{RB}(f_{0})$] for the pixel of interest that are given by the following expressions:
\begin{equation}
WCP_{RB}(f_{0}) = G(f_{0}, \delta f) \|CS_{RB}(f)\|
\end{equation}
\begin{equation}
\phi_{RB}(f_{0}) = G(f_{0}, \delta f) \tan^{-1} \left( \frac{ Im \{C_{RB}(f)\} } { Re \{C_{RB}(f)\} } \right)
\end{equation}
\begin{equation}
C_{RB}(f_{0}) = G(f_{0}, \delta f) \left( \frac{ CS_{RB}(f)^2 } { CS_{RR}(f) * CS_{BB}(f) } \right)
\end{equation}
\begin{equation}
T_{RB}(f_{0}) = \frac{ \phi_{RB}(f_{0}) }{ 2 \pi f_{0} }
\end{equation}
Repeating this step for all of the neighboring pixels results in the construction of 10$\times$10 pixel maps of the three quantities for each filter frequency, $f_{0}$.

The panels in Figure~\pref{fig1} show coherence, cross-power, phase and phase travel-time maps from the spectral analysis of one pixel (coordinates 181, 97) in the \trace{} 171\AA{} dataset from 1 July 1998 12:03, \-- 1 July 1998 13:02 UTC (see below) at a filter frequency of $f_{0}$ = 5.5mHz. In particular, we see that  the coherence map (panel A), like that shown in the approach of \citet{2007Sci...317.1192T}, has a region, or ``island'', where the spectral coherence is large ($>$0.6) at this frequency across numerous neighboring pixels. It is the properties of the strong coherence island and its mappings in phase-time and cross-power that form the cornerstone of the analysis presented. They can provide a quick and reliable estimate of the propagation direction and phase-speed of the disturbance present, from which reasonable errors can be determined. 

After isolating the relevant high coherence pixels in the phase-time diagnostic map, we compute a distance-minimizing linear fit\footnote{In this case we seek the best fitting straight line through the cluster of points in the island that also minimizes the distance between the points and that straight line \citep[see, {\it e.g.}][]{2003drea.book.....B}. We have found that this approach is not subject to singularities for clusters of points that are orthogonal, or nearly orthogonal, to the horizontal as is the case in standard least-squares fits when the dependent and independent variables are interchangeable or both have measurement errors.} to the cluster of points in the island and calculate a distance d along the resulting straight line as well as an angle (and the error on the angle) at which the disturbance is moving (relative to North \-- South in the coherence map). We also use the cluster of pixels to extract a measure of two coherent length-scales, one that is parallel to the measured propagating direction (the coherence length), and one that is perpendicular to it (the coherence width). These last diagnostics may contain physical information about the process exciting the disturbance and so we retain them at little computational cost.

We then compute the phase-speed of the disturbance and its associated error by forming a scatter plot of distance $d$ (from the reference pixel) versus measured phase-time (see, {\it e.g.}, Figure~\pref{fig2}). The gradient of the least-squares linear fit to these data provides an estimate of the phase speed of the disturbance, errors in the gradient are determined by assessing a range of possible values that can achieve a similar quality of fit to the data \citep[see, {\it e.g.} {\it Sect.}~6 of][]{2003drea.book.....B}. We repeat the previous step for each filter required unless there are fewer than 6 pixels in the island. 

This process is then repeated for all of the other pixels in the original image with the result that we develop images of the apparent phase-speed, angle, (their respective errors) and additional measures of the island-mean coherence and cross-power (see, e.g, Figure~\pref{fig5a} and {\it Sect.}~\pref{s:results}).

\subsection{Temporal Domain Analysis}\label{s:ptr}

At the cost of a considerable increase in computation time, we can substitute the phase-time and phase-speed measurements by performing their calculation in the temporal domain. This step involves the cross-correlation of the Fourier filtered timeseries for all of the pixels in the high coherence island relative to the reference pixel. The cross-correlation function at each neighboring pixel is a Gabor wavelet that, when fitted, yields information about the phase (and energy carrying group) travel-times of the disturbance \citep[][]{2004ApJ...613L.185F}. In practice we cannot execute the temporal domain analysis for \sdo/AIA as the computational cost is prohibitive for a method that is essentially mathematically equivalent to the one presented above. However, we do note that the process is illustrative of the general technique, the quantities and measures developed for the analysis and so we have chosen to provide an example.

\begin{figure} 
\centerline{\includegraphics[width=0.5\textwidth,clip=true]{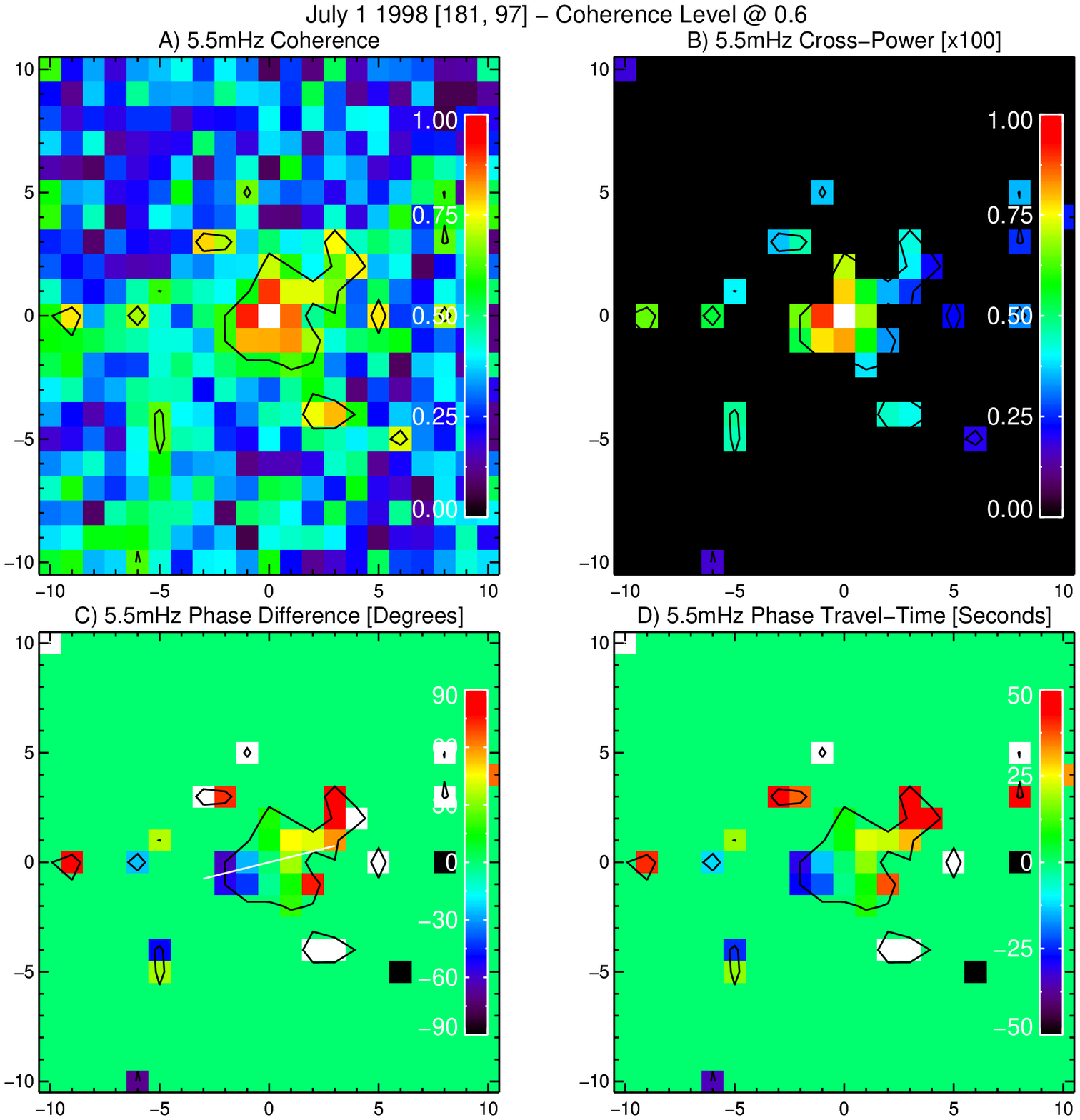}
                  \includegraphics[width=0.5\textwidth,clip=true]{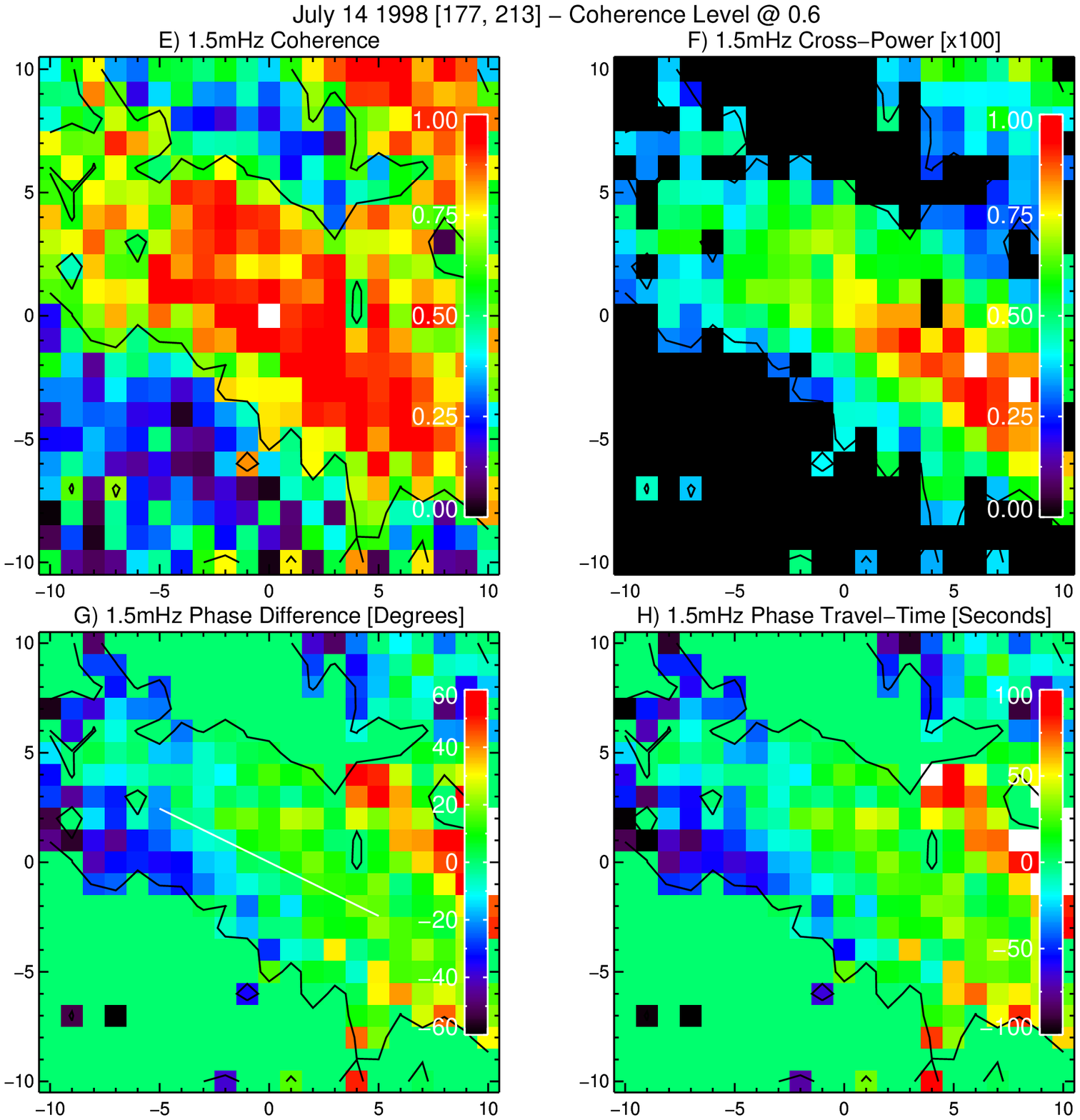}}
\caption{Examples of the coherence, weighted cross-power, phase difference and phase travel-time for for the 1 July 1998 (panels A-D) and 14 July 1998 (panels E -- H) datasets for 5.5~mHz and 1.5~mHz Fourier filter samples respectively. In each case the reference pixel is at 0,0, the solid black contour outlines regions of highly coherent signal and the fitted angle (solid white lines in panels C and G) to central coherence ``island'' is shown as a solid white line in panels C (14.6$^\circ$) and G (-30.4$^\circ$).}\label{fig1}
\end{figure}

\begin{figure} 
\centerline{\includegraphics[width=0.5\textwidth,clip=true]{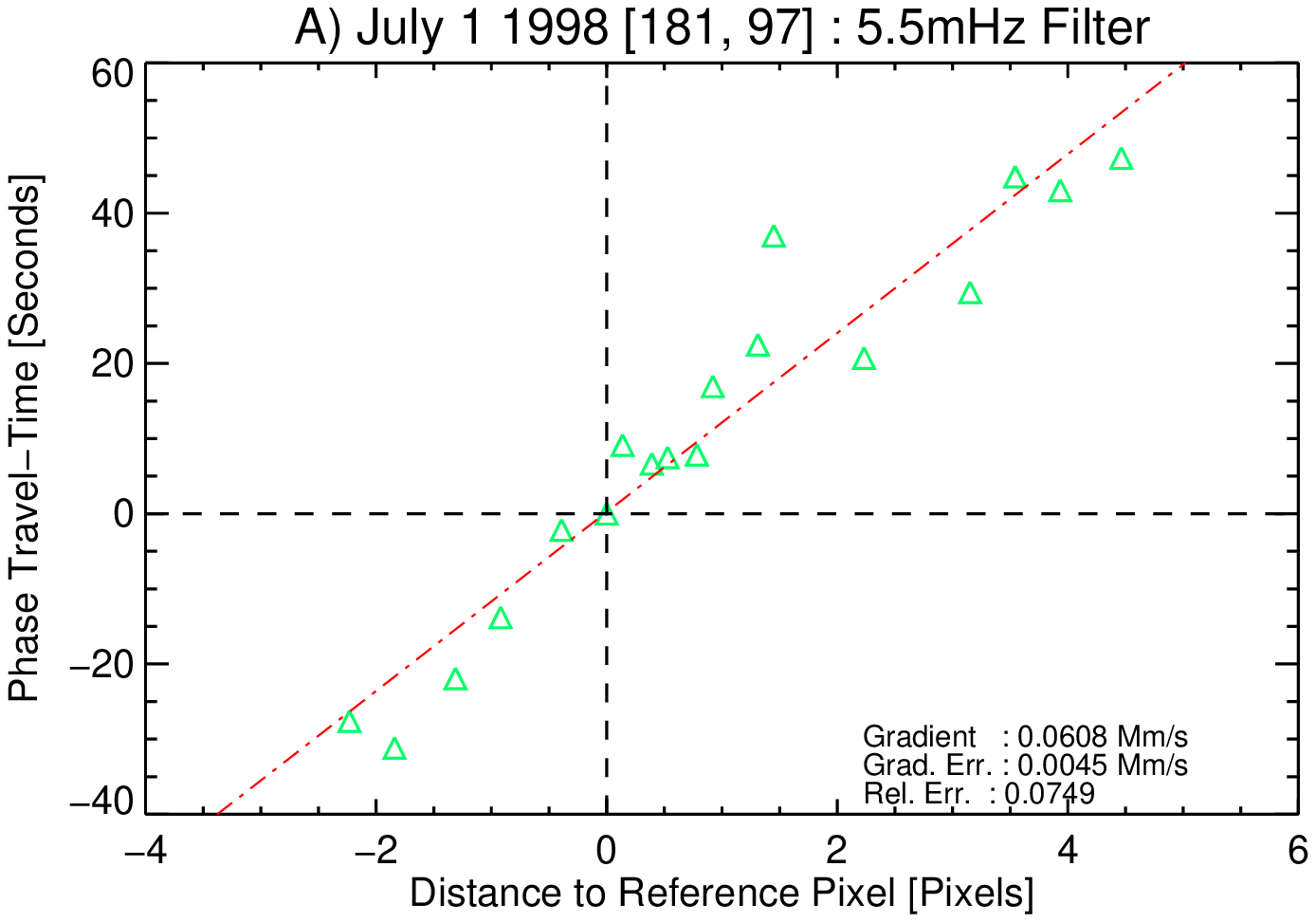}
                  \includegraphics[width=0.5\textwidth,clip=true]{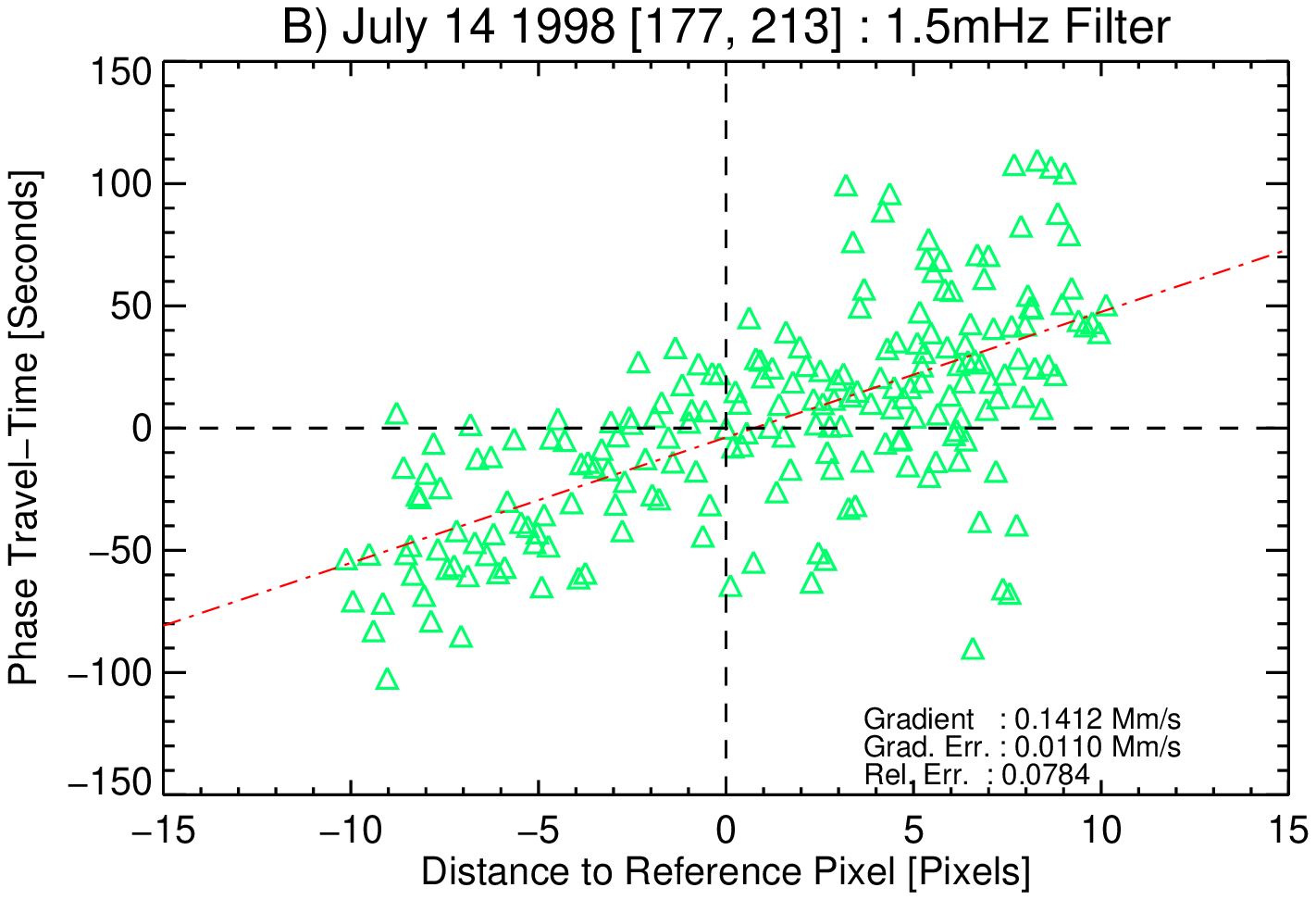}}
\caption{Examples of phase speed determination for the cases shown in Figure~\pref{fig1}. In each case the green triangles mark the positions from the reference pixel and measured phase travel-time at that pixel. The gradient of the least-squares linear fit (red dot-dashed line) yields the phase-speed of the propagating signal.}\label{fig2}
\end{figure}

\begin{figure} 
\centerline{\includegraphics[width=0.75\textwidth,clip=true]{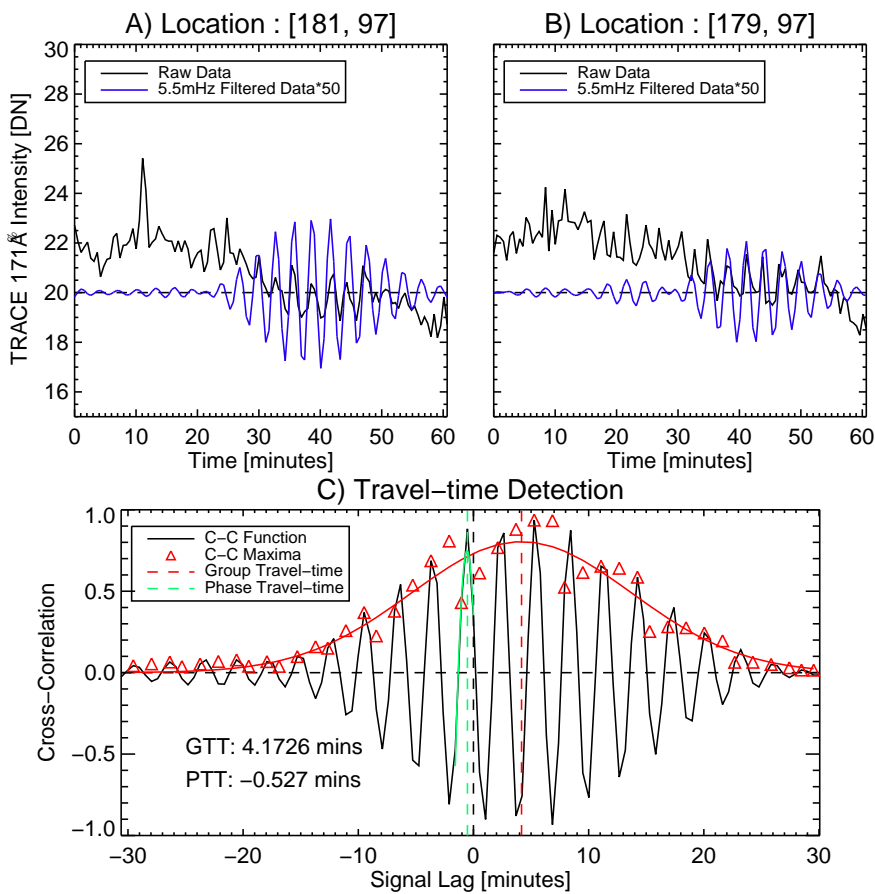}}
\caption{Example pixel-to-pixel travel-time analysis in the temporal domain. In panels A and B we show the \trace{} 171\AA{} lightcurve (black solid line) and its 5.5~mHz filtered counterpart (blue solid line) between the reference pixel [181, 97] and pixel [179, 97] for the 1 July 1998 dataset ({\it cf}. Figure~\pref{fig1}). In panel C we show the derived cross-correlation function (black solid line), envelope maxima (red triangles) and fit to the central peak (green solid line). The vertical dashed lines in panel C show the positions of the estimated group travel-time (red; -4.17 minutes) and phase travel-time (green; -0.527 minutes, {\it i.e.}, consistent with the values in Figures~\pref{fig1} and propagating left to right which is consistent with the movie of Figure~\pref{fig4}) respectively.}\label{fig3}
\end{figure}

\begin{figure} 
\centerline{\includegraphics[width=0.5\textwidth,clip=true]{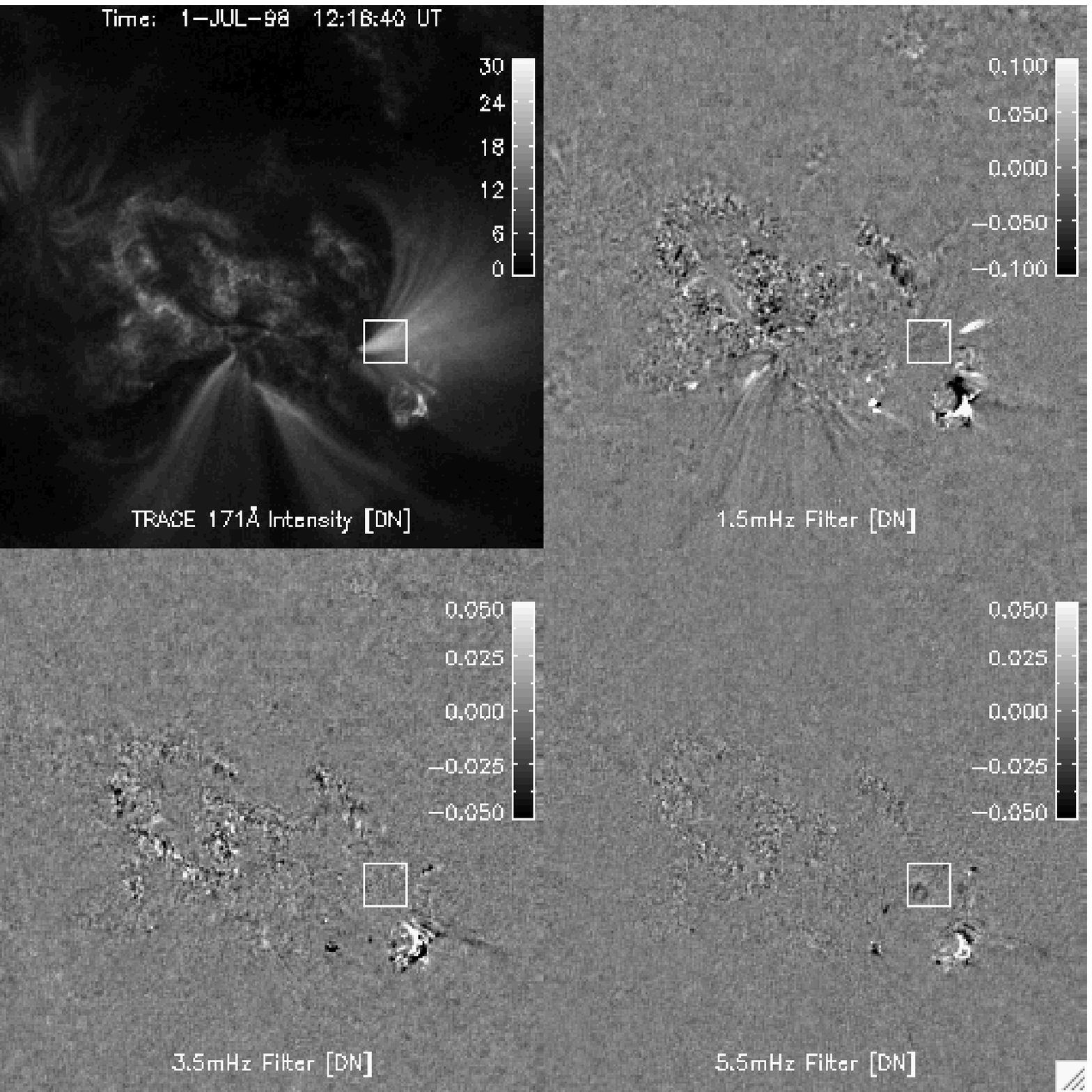}
                  \includegraphics[width=0.5\textwidth,clip=true]{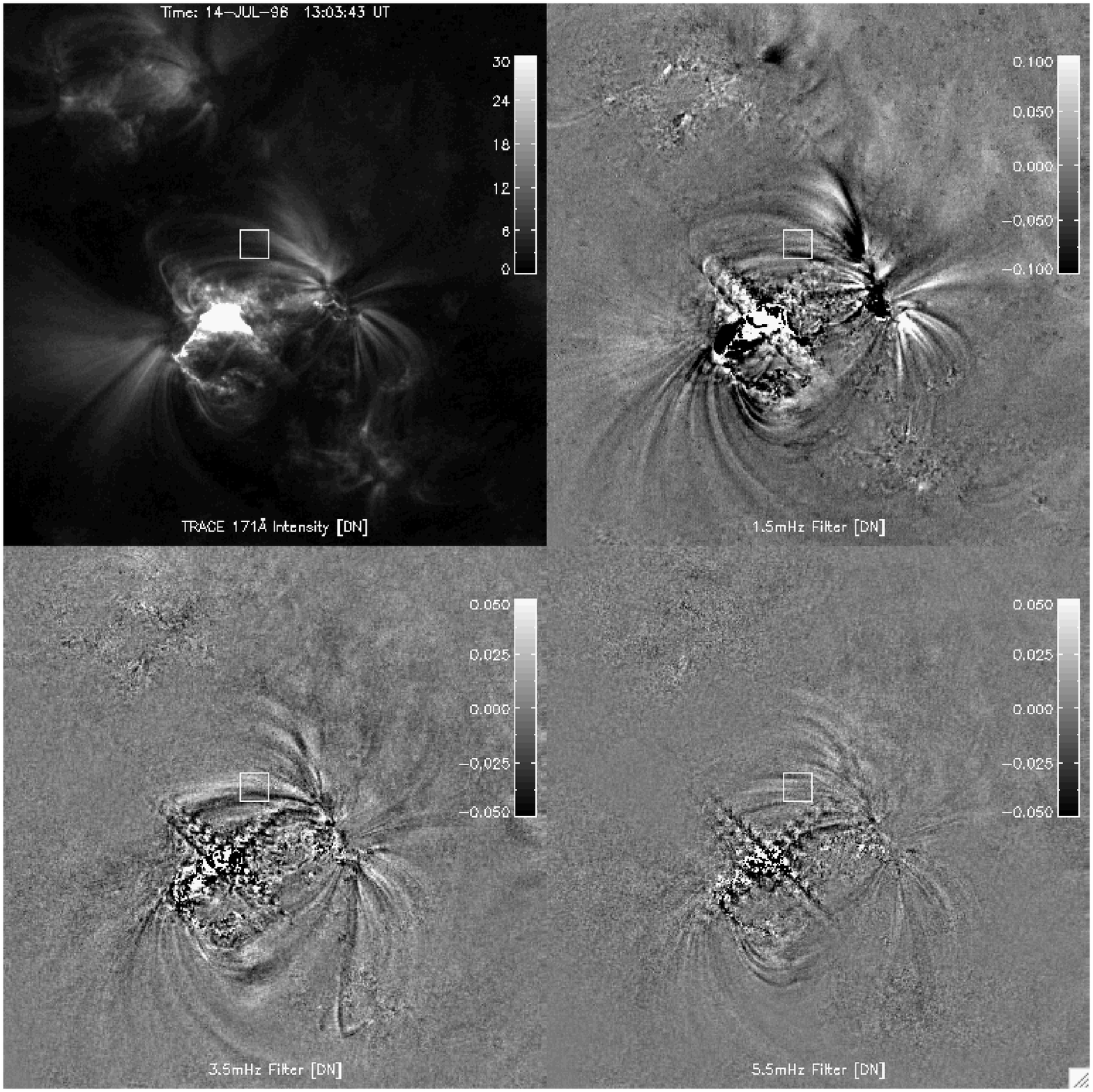}}
\caption{Illustrations of the two \trace{} 171\AA{} timeseries used in the presented analysis; 1 July 1998 (left) and 14 July 1998 (right). In each panel we show (clockwise from the top left) snapshots of the \trace{} 171\AA{} intensity, 1.5~mHz, 5.5~mHz, and 3.5~mHz filtered timeseries respectively. The boxes marked on the panels correspond to the regions used to construct Figure~\pref{fig1} and the subsequent analyses. See the online edition of the Journal for animations of these figures.}\label{fig4}
\end{figure}

An example of Òtwo-stepÓ\footnote{Typically we would fit both phase and group information simultaneously, as is done in \citet{2004ApJ...613L.185F} and the other references provided above, but for the data at hand, where often the group envelope is not well defined, we fit them separately.} cross-correlation fitting is shown in Figure~\pref{fig3}. First, we fit the envelope of the function to extract group information, after which we fit the peak of the cross-correlation nearest to the lag origin to obtain phase information. The solid black line shows the cross-correlation of the reference timeseries used to create Figure~\pref{fig1} and that from another pixel (coordinate [181, 91] in the figure) as a function of lag in seconds. The group behavior is determined by fitting a Gaussian (dot-dashed line) to the envelope of the cross-correlation function that is outlined by the absolute values of the function maxima (red triangles) \-- the center of the fitted Gaussian measures the group travel-time (or energy transport time). The phase travel-time can then be extracted by fitting the peak of the cross-correlation function nearest to the origin. Using either a parabolic of Gaussian fit we can extract phase-times (the center of the Gaussian or parabolic fit) down to an accuracy of about one tenth of the image cadence. For the case considered the vertical dashed lines in panel C of Figure~\pref{fig3} show the positions of the estimated group travel-time (red; -4.17 minutes) and phase travel-time (green; -0.527 minutes. For good measure, we note that the latter value is consistent with the values in Figures~\pref{fig1} and demonstrates that the disturbance is propagating left to right which is also consistent with the movie of Figure~\pref{fig4}) respectively.

Repeating the steps used in the spectral analysis to compute the phase-speed we perform a least-squares linear-fit to the scatter plot of pixel distance and measured phase-travel times. We note that it is also possible to investigate the group-speed of the disturbance - by simply replacing the phase travel-times with the measured group travel-times and repeating the fit to the scatter. 

\subsection{Improving Algorithm Efficiency}\label{s:effic}

The algorithm described in Sects.~\pref{s:sda} and~\pref{s:ptr} is quite slow, since it calculates cross-correlations or cross-spectral properties for 100 neighboring points at each location of  the datacube. We have improved the efficiency of the algorithm by a large factor (typically of order 20) in the following way. For each location, we define a map of locations for which cross-correlations with the central pixel [$x_{0},y_{0}$] need to be calculated. At the first iteration this map contains only the 4 immediate neighbors of the central pixel. We then determine which of these locations has a coherence larger than the cutoff (0.6 in our case). If none of these locations has a coherence larger then the cutoff, the algorithm ends the calculations for pixel [$x_{0},y_{0}$], and moves on to [$x_{0}+1,y_{0}$]. If one or more locations shows a coherence larger than the cutoff, we repeat the iteration by calculating a new map of points for which cross-correlations need to be calculated. This map will now contain the immediate neighbors of the high coherence locations calculated in the previous round. We exclude from this map all locations that have already been calculated. After this the algorithm again checks whether any of the new points has a coherence larger than the cutoff, continuing until no new locations are above the cutoff. In this fashion we can usually reduce the number of calculations from 100 to around 5, since many locations show little coherence with their neighbors, and only a select few locations show significant coherence with their neighbors. This increase in efficiency has a critical impact on the usefulness of the algorithm for searching through large datasets, such as those that are expected from \sdo/AIA.

\section{Test Data}\label{s:data}

To illustrate the algorithm we have applied it on two different \trace{} datasets. Dataset I was taken on 1 July 1998 from 12:03 to 13:02 UTC in 171\AA{}, with a cadence of 31.5s. Dataset II was taken on 14 July 1998 from 12:45 to 13:42 UTC in 171\AA{} during the so-called Bastille day flare and has a cadence of 72.3 seconds. The first dataset contains an active region with some  slow-mode running waves in one of the loops associated with a sunspot. The second dataset has been the subject of several papers \citep[see, {\it e.g.}][]{1999ApJ...520..880A, 2002SoPh..206...99A} and contains the ``classical example'' of transverse oscillations. Before applying the algorithm, we performed the calibration steps included in {\texttt trace\_prep.pro} (as part of the \trace{} tree of solarsoft IDL), which includes the subtraction of dark current/pedestal, correction for the ccd gain and lumogen degradation, correction of bad pixels and normalization of all images to remove differences in exposure time. In addition, we performed several iterations of cosmic ray removal using the {\texttt trace\_unspike.pro} routine which removes excessively bright spikes by comparing the data with previous and following timesteps and replacing the spike with a temporal and spatial average of the surrounding pixels. As a final step we derotate the data by using {\texttt poly\_2d.pro} to perform sub-pixels shifts that are calculated using {\texttt diff\_rot.pro}. We also perform a 2$\times$2 rebinning to reduce the \trace{} data to a spatial resolution of 1 arcsecond. We have tested the impact of the rebinning extensively and found that rebinning to the spatial resolution does not lead to any loss of useful information about the oscillations, and in fact improves the signal to noise of the sometimes weak oscillatory signals. Of course, rebinning the data naturally speeds up the algorithm by a factor of 4 compared to running it on the data at its original resolution.

Figure~\pref{fig4} show snapshots of the raw and Fourier filtered timeseries of the 1 July 1998 (left) and July 14 1998 (right) datasets respectively. In each case, the upper left panel shows a snapshot of the intensity timeseries while other panels show the filtered timeseries with a Gaussian filter centered around 1.5~mHz, 3.5~mHz and 5.5~mHz respectively. In the left panel we see that the active region contains several sets of loops, with the right-most fan associated with a sunspot being highlighted in a box. The loops in the southern part of the active region are associated with plage regions. We see that the sunspot fan contains a clear 180s signal which is visible as a dark spot in the 5.5~mHz snapshot (lower right) and an oscillatory signal in the same panel of the provided movie. These kinds of oscillations have been described extensively in a series of papers by \citet{2002SoPh..209...61D, 2002SoPh..209...89D}. In addition, the moss regions towards the north and middle of the active region show a predominance of 1.5 and 3.5~mHz power. This is not surprising and has been noted before by \citet{2003ApJ...590..502D} who showed that the periodic obscuration of upper transition region emission by chromospheric jets plays a dominant role in the EUV (semi-)periodicities. A very intriguing finding is that there is significant power in the 1.5~mHz filtered movies and snapshot in the plage-related loops towards the south of the active region. We will return to these low frequency oscillations in Sects.~\pref{s:results} and~\pref{s:discuss}.

As a cautionary note we see that generally, these filtered movies show not only periodic signals, but any signal that shows power at the chosen frequency. Because of the nature of the Fourier transform, this includes signals that are caused by abrupt changes in the brightness, such as cosmic rays or sudden loop brightenings ({\it e.g.}, in the case of emerging flux reconnecting with overlying field, or in post-flare loop arcades). To exclude some of those signals when we run our algorithm on the data, we take care to use a Hanning window (reducing the influence of trends), and apply a mask that excludes from our calculations all pixel locations whose raw time series show deviations that are more than 3.5 times the standard deviation of the time series. This practically excludes many of the locations with cosmic rays and sudden brightenings.

This approach is especially useful in the case of a flare where many locations show significant and sudden brightenings, such as is the case in dataset II, and shown in the right panel. The filtered movies and snapshots show many locations with significant oscillatory power, which is borne out by the unfiltered time series. The flare is associated with a coronal dimming, which leads to significant ``wiggling'' of the coronal loop structures, especially on the 5-10 minute time scales. Very prominent in these snapshots are also the post-flare loops in the middle of the active region. Most interesting is that the transversely oscillating loops show up very clearly in the movies and snapshots, especially at 3.5~mHz (lower left) and to a lesser degree at 5.5~mHz (lower right).

\section{Results}\label{s:results}

\subsection{Interpretation of Raw Results}\label{s:raw}

We present results of the spectral algorithm for both datasets. As we have stated above the results for the temporal code are very similar but run at a much lower speed so that, as such, it is not adequate for our ultimate goal of applying this technique to \sdo/AIA. In principle our algorithm produces for each location, and for each frequency we choose to filter on, the following average properties of the island of high coherence around each location: cross-spectral power, average coherence, phase speed of propagating signal, error on the phase speed, the angle of propagation of the disturbance, the correlation length and the correlation width. To reduce the amount of noise in the resulting maps of the active region, we reject locations for which the island of high coherence consists of fewer than 6 neighbors. 

\begin{figure} 
\centerline{\includegraphics[width=0.75\textwidth,clip=true]{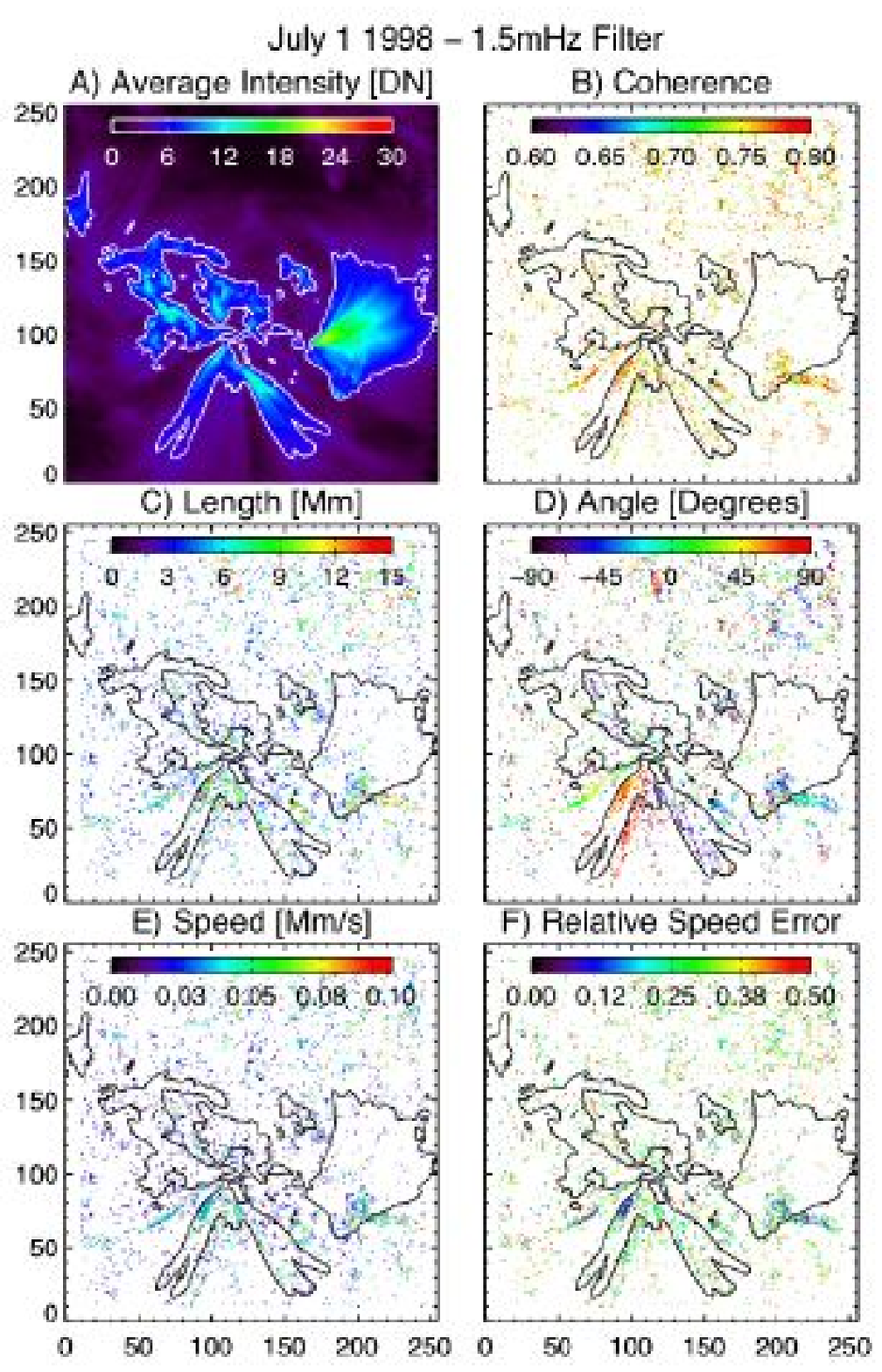}}
\caption{Results of the coronal wave detection algorithm for the 1.5~mHz filtered timeseries for the 1 July 1998 dataset. We show the average \trace{} 171\AA{} intensity image, weighted signal coherence, length, width, angle, phase speed, and relative error of the phase speed computed from the large coherence island.}\label{fig5a}
\end{figure}

\begin{figure} 
\centerline{\includegraphics[width=0.75\textwidth,clip=true]{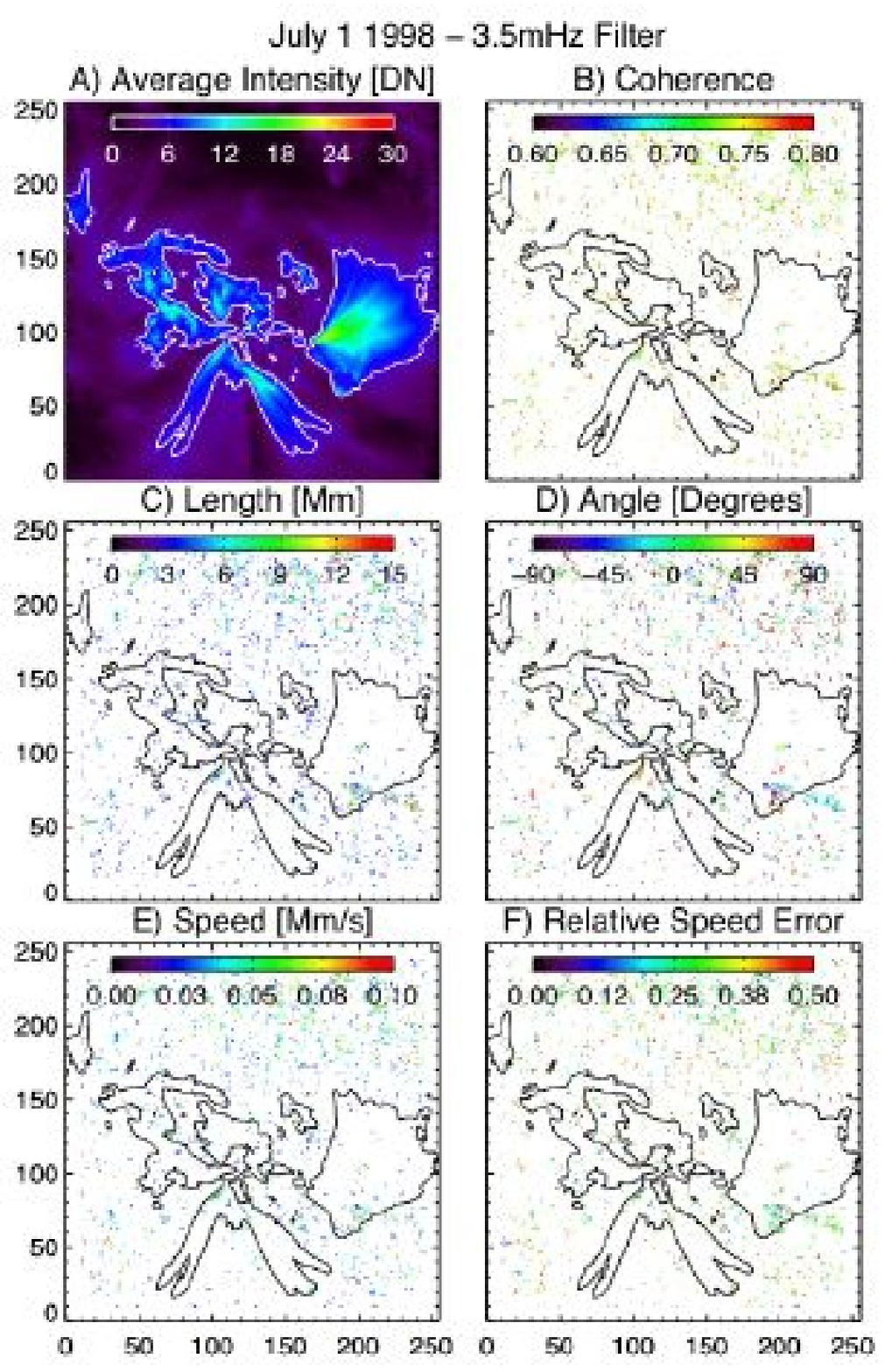}}
\caption{Results of the coronal wave detection algorithm for the 3.5~mHz filtered timeseries for the 1 July 1998 dataset. We show the average \trace{} 171\AA{} intensity image, weighted signal coherence, length, width, angle, phase speed, and relative error of the phase speed computed from the large coherence island.}\label{fig5b}
\end{figure}

\begin{figure} 
\centerline{\includegraphics[width=0.75\textwidth,clip=true]{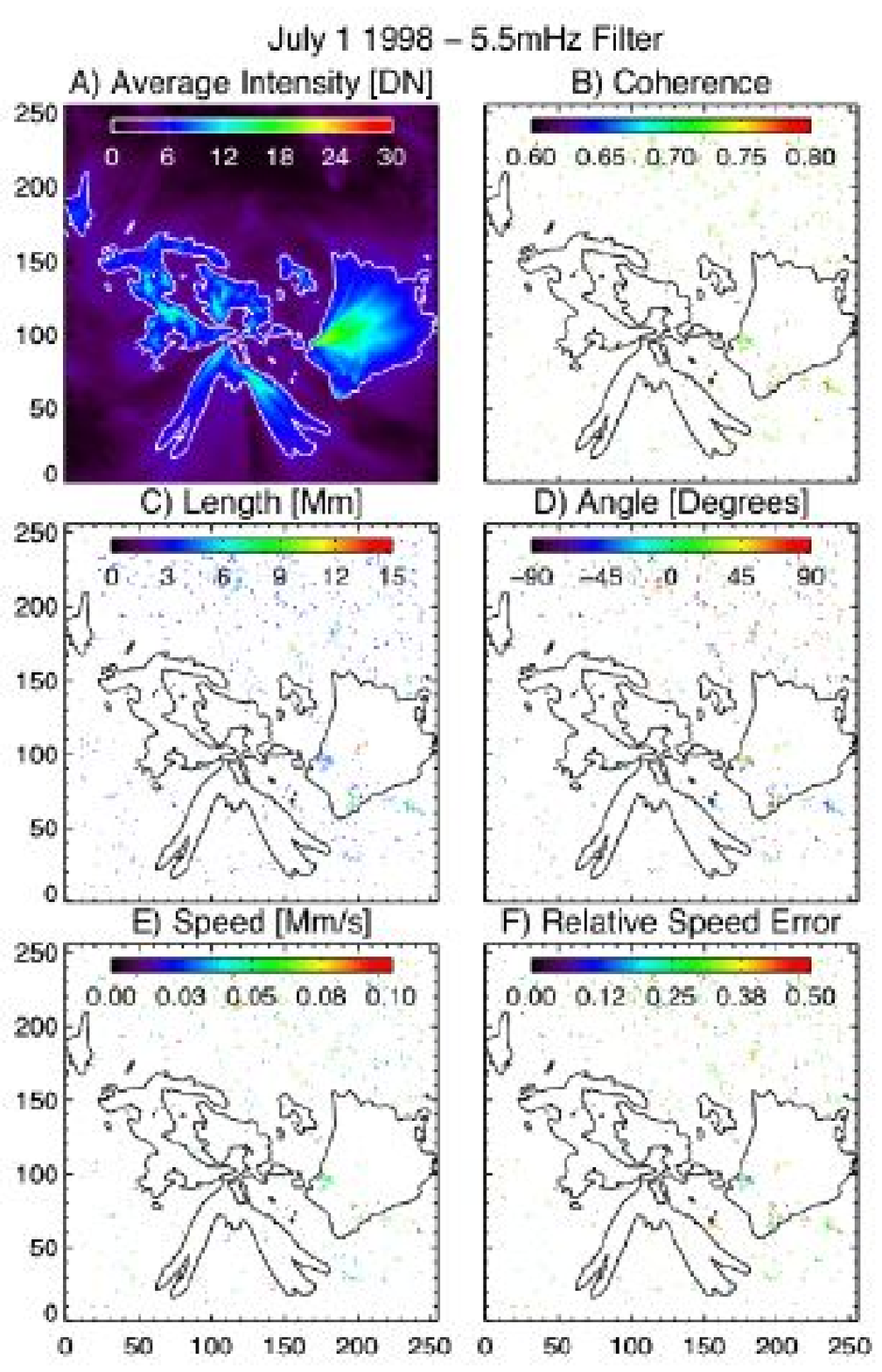}}
\caption{Results of the coronal wave detection algorithm for the 5.5~mHz filtered timeseries for the 1 July 1998 dataset. We show the average \trace{} 171\AA{} intensity image, weighted signal coherence, length, width, angle, phase speed, and relative error of the phase speed computed from the large coherence island.}\label{fig5c}
\end{figure}

\begin{figure} 
\centerline{\includegraphics[width=0.75\textwidth,clip=true]{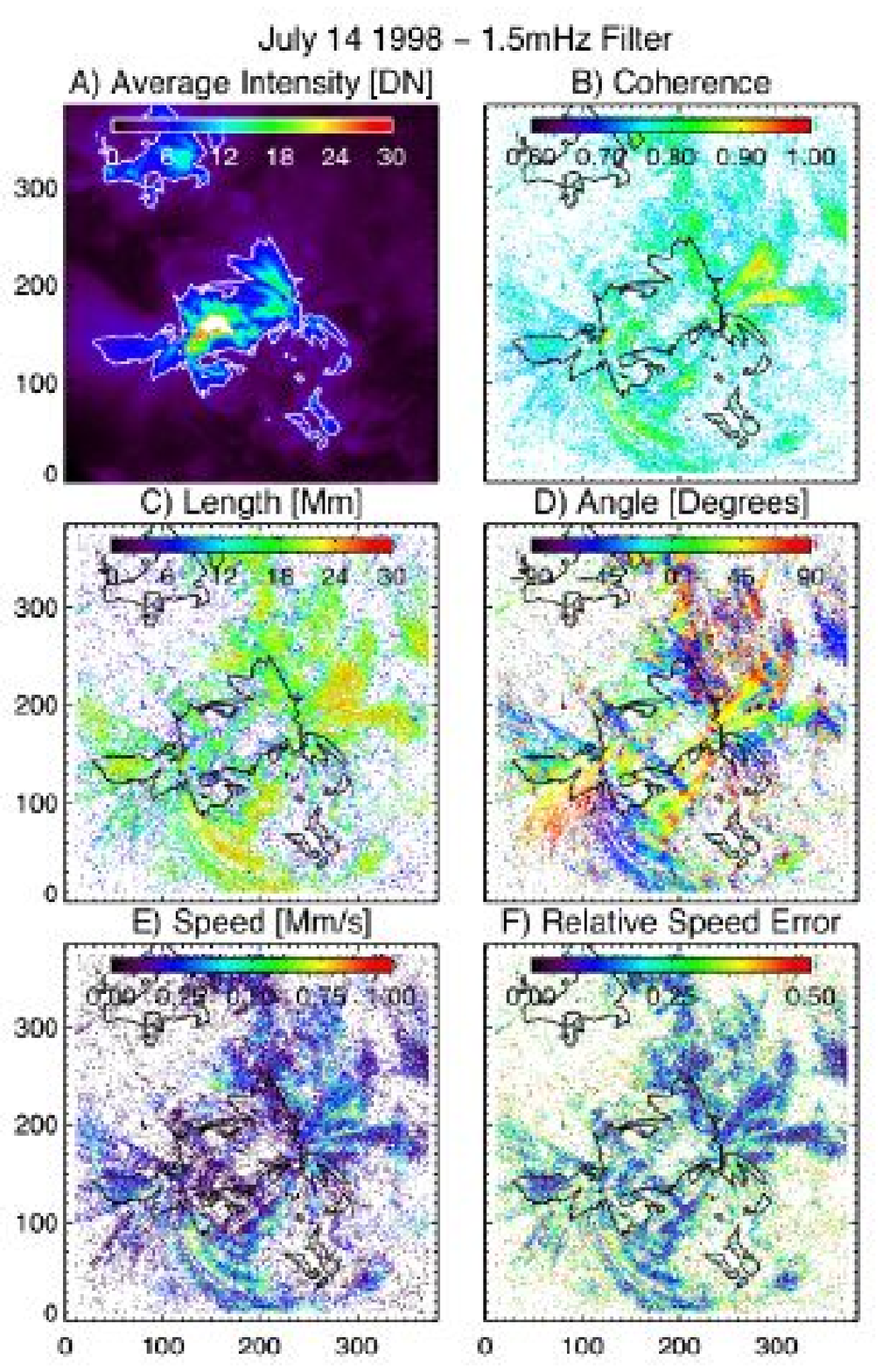}}9
\caption{Results of the coronal wave detection algorithm for the 1.5~mHz filtered timeseries for the 1 July 1998 dataset. We show the average \trace{} 171\AA{} intensity image, weighted signal coherence, length, width, angle, phase speed, and relative error of the phase speed computed from the large coherence island.}\label{fig6a}
\end{figure}

\begin{figure} 
\centerline{\includegraphics[width=0.75\textwidth,clip=true]{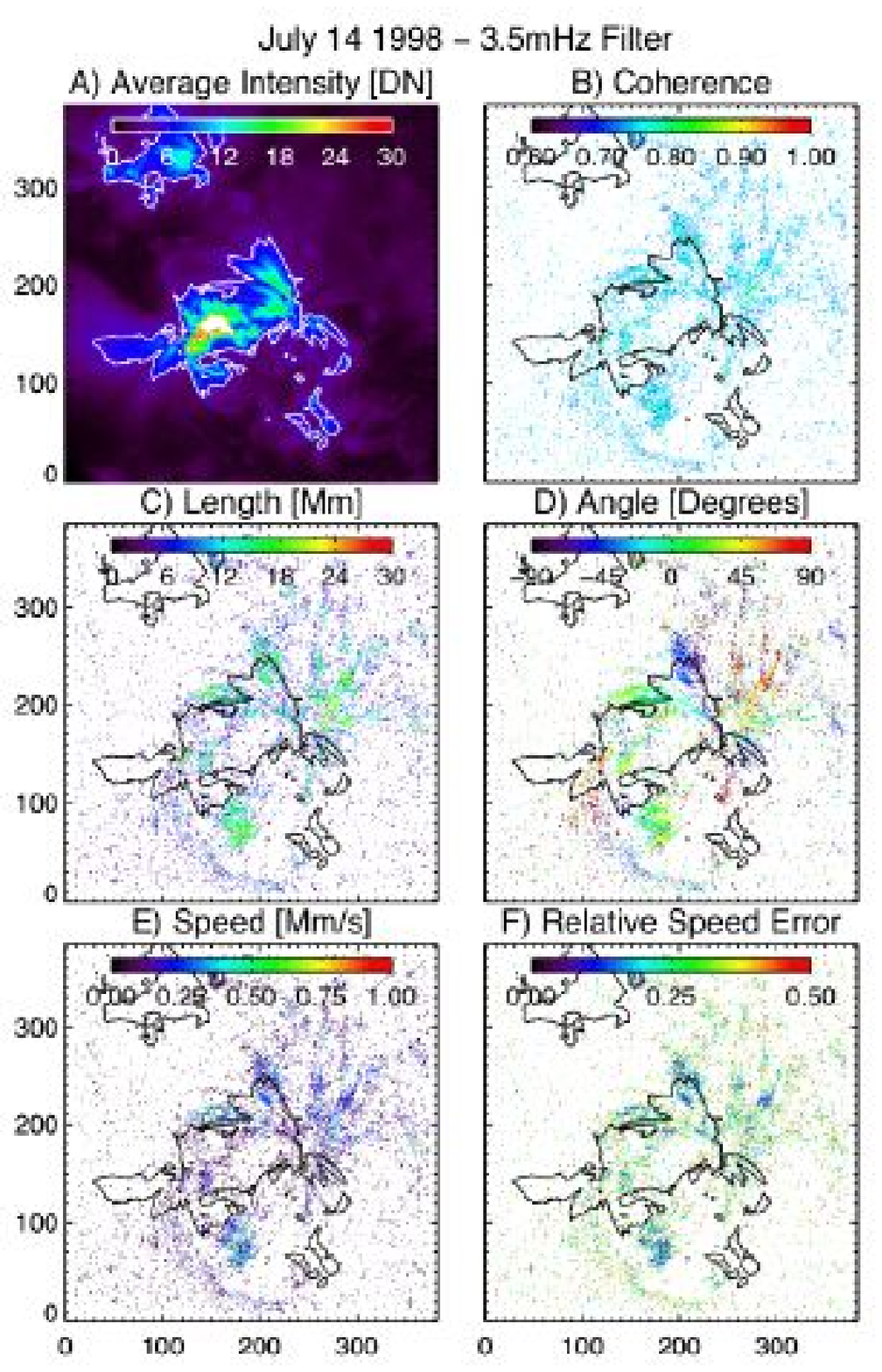}}
\caption{Results of the coronal wave detection algorithm for the 3.5~mHz filtered timeseries for the 14 July 1998 dataset. We show the average \trace{} 171\AA{} intensity image, weighted signal coherence, length, width, angle, phase speed, and relative error of the phase speed computed from the large coherence island.}\label{fig6b}
\end{figure}

\begin{figure} 
\centerline{\includegraphics[width=0.75\textwidth,clip=true]{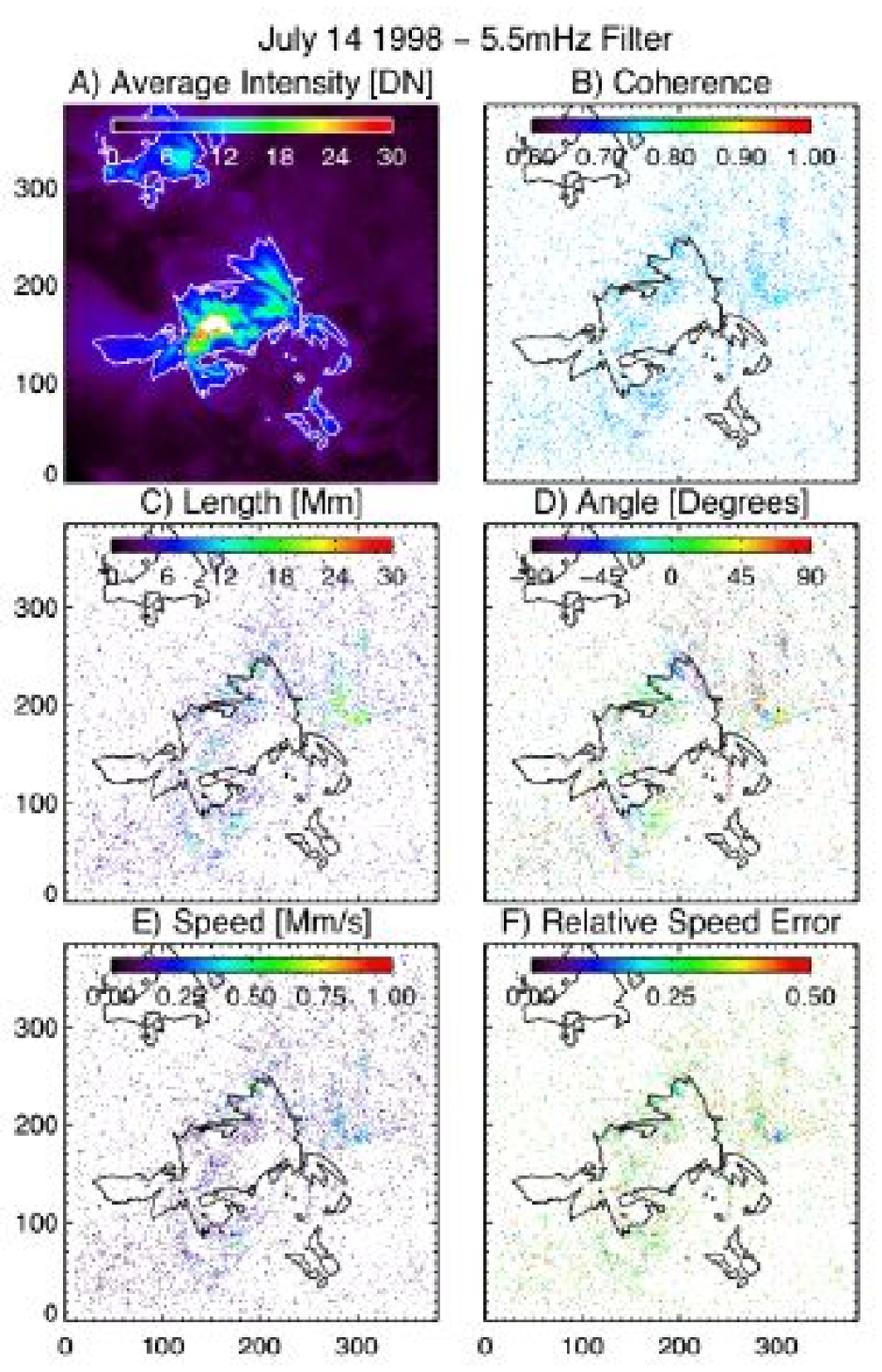}}
\caption{Results of the coronal wave detection algorithm for the 5.5~mHz filtered timeseries for the 14 July 1998 dataset. We show the average \trace{} 171\AA{} intensity image, weighted signal coherence, length, width, angle, phase speed, and relative error of the phase speed computed from the large coherence island.}\label{fig6c}
\end{figure}

The resulting maps of the properties described above show only locations where the signal is significantly correlated with at least 5 neighbors (Figures~\pref{fig5a}, ~\pref{fig5b} and~\pref{fig5c} for the frequencies 1.5, 3.5 and 5.5~mHz for dataset I, and Figures~\pref{fig6a}, ~\pref{fig6b} and~\pref{fig6c} for dataset II). What is striking in these maps is the large amount of locations that have (oscillatory) correlation with their neighbors. However, since our algorithm is so sensitive, many of these locations actually show ``false positives'' in which a very small (mostly instrumental or spurious) oscillatory signal is correlated with a few neighboring pixels. Nevertheless, some interesting physical results immediately show up that confirm the findings of our cursory analysis of the Fourier-filtered time series in Section 3. For example, the 3.5~mHz map of phase speeds of dataset I (Figure~\pref{fig5b}) is dominated by locations where moss occurs, as can be expected from previous analyses \citep[{\it e.g.},][]{2003ApJ...590..502D}. The phase speeds are quite variable and of order 10-50 km~s$^{-1}$ in the moss regions, which is perhaps not surprising since the correlations with neighboring pixels are most likely caused by chromospheric jets (with similar velocities to those we find; \citep[{\it e.g.},][]{2003ApJ...590..502D, 2006ApJ...647L..73H} moving across TRACE pixels. The 3.5~mHz map also shows evidence of oscillatory power in the bottom of some coronal loops ({\it e.g.}, around 110, 90) with phase speeds of order 50 \-- 100~km~s$^{-1}$. 

The 5.5~mHz map of dataset I (Figure~\pref{fig5c}) is much more sparsely populated with only a few major concentrations of significant oscillatory power towards the lower right of the map. These locations are exactly where we find (through visual examination) oscillations in the original and Fourier-filtered time series: our algorithm nicely locates the sunspot oscillations at 180 second periods (around 180, 90). The phase speeds here are of order 50\--150~km~s$^{-1}$, which is within the range that is expected for slow-mode propagating disturbances \citep[][]{2002SoPh..209...61D,2002SoPh..209...89D}. 

The 1.5~mHz results (Figure~\pref{fig5a}) show large parts of the southern coronal loops with significant oscillatory disturbances that propagate along the loops with phase speeds that are similar and of order 50 \-- 150~km~s$^{-1}$. If these oscillations are real (see {\it Sect.}~\pref{s:summary}), they are interesting, since they are of much longer periods than have been previously reported (typically five minutes in plage-related loops). We have verified that the signals are not introduced by slow pointing drifts on timescales of ten minutes by performing the same analysis using data that is co-aligned with various co-alignment strategies (see {\it Sect.}~\pref{s:summary}). In addition, co-alignment issues would lead to similar signals in similarly oriented loops, {\it e.g.}, the loops at the northern edge of the sunspot fan. Our algorithm does not find any sign of significant propagating and oscillatory disturbances at 1.5~mHz in those loops.

The results of dataset II are similarly interesting. First of all, the phase speeds shown in all frequencies are systematically higher in this dataset, with most being of order 150 \-- 500~km~s$^{-1}$. This is compatible with the fact that this dataset is dominated by a flare, coronal dimming and the associated (mostly fast magneto-acoustic mode) waves and/or standing waves (transverse oscillations) that have been well documented in previous work focusing on the Bastille day flare \citep[{\it e.g.},][]{1999ApJ...520..880A, 2002SoPh..206...99A}. The 3.5~mHz maps (Figure~\pref{fig6b}) of phase speed are dominated by locations of significant oscillatory power at the same locations where previous analysis (based on visual inspection) had found evidence of transversely oscillating loops ({\it e.g.}, around 170, 200 or 180, 70 or 240, 110). In addition, our algorithm finds a range of locations with oscillatory power that have not been discussed previously, such as the western side of the active region (the general area from {\it x} $>$ 200 and {\it y} $>$ 200). The phase speeds in the 3.5~mHz maps are of the order of several hundred km~s$^{-1}$, which is to be expected for standing waves or propagating fast-mode waves in the corona.

The 5.5~mHz maps (Figure~\pref{fig6c}) are again much more sparsely populated, with only a small subset of locations showing significant oscillatory power. This is perhaps not surprising since previous analyses have shown that this particular flare led to transverse oscillations with periods of order five minutes. However, the fact that we do find power at 5.5~mHz ($\approx$three minute periods), and that it occurs at mostly the same locations as in the 3.5~mHz maps may suggest that the transverse oscillations are not dominated by a single peak in the Fourier spectrum, but rather have a range of periods. 

The 1.5~mHz maps (Figure~\pref{fig6a}) are very intriguing with most of the dimming regions showing significant oscillatory power that propagates at phase speeds of order 100-500~km~s$^{-1}$. However, in this case it is more difficult to clearly distinguish whether the resulting phase speeds and locations are uniquely the sites of an oscillatory disturbance propagating, or whether the sudden coronal dimming triggered by the flare is partially responsible for the results. However, visual inspection of the Fourier-filtered time series suggests that the coronal dimming is indeed accompanied by transient and quasi-periodic ``wiggling'' of at least some of the coronal structures in these regions. The observed range of phase speeds may then well be the propagation speed of the fast mode waves that are associated with such a drastic reorganization of the corona.

In the above we have focused on the locations of significant oscillatory power and the associated phase speeds of the observed oscillatory disturbances. Our algorithm also naturally provides the direction in which the propagation occurs, as well as the coherence length and width of the islands of high coherence. The observed propagation direction usually is aligned with the dominant intensity structure (coronal loop or fan) that carries the disturbance. The coherence width and length are measures of how large the coherence island is, with the width by definition shorter than the length, and the difference between both larger as the island of high coherence is more oblong in shape. Typical coherence lengths are of order three \-- six~Mm in dataset I, with longer coherence lengths for lower frequencies. For dataset II coherence lengths are of similar size, except for the low frequency results which have coherence lengths that are up to 10 \-- 35~Mm in the regions where the coronal dimming dominates. This is not surprising, since the dimming impacts a large area of the active region.

\subsection{Reducing False Positives}\label{s:rfp}

As we mentioned in the above, our algorithm is so sensitive that it finds a significant number of false positives \-- locations where instrumental or cosmic ray artifacts lead to spurious coherent signals. In addition, many of the locations have values for the phase speed that are based on poor fits between the distance along the propagation direction and the travel time, which can lead to significant  errors on the parameters that are calculated. While the value of the maps we calculate is clear, the required human interpretation of what constitutes a false positive would render an automated classification (which is crucial for \sdo/AIA) difficult. Fortunately we have found a method (which we describe in the following) to significantly reduce the number of false positives.  

We take several steps to prune the results down to locations where the observed signal is most likely caused by a real coronal signal, as opposed to an instrumental or detection artifact. We start with a map that contains locations that have enough neighbors (as defined in {\it Sect.}~\pref{s:method}) with high coherence. As a first step to winnow down the number of false positives, we use the error on the determination of the phase speed (from the least-squares fit of distance versus travel-times). We reject all locations for which the relative error on the phase speed is larger than 0.3 ({\it i.e.}, the error divided by the phase speed). This removes from our map a large number of locations where the travel times are not well correlated with distance along the island of  high coherence. Such a poor correlation leads to a poor linear fit and a large error on the phase speed. To reduce the large number of locations caused by instrumental artifacts we then reject all locations in which the average brightness is below a certain threshold (1~DN~s$^{-1}$). This removes a large number of artifacts that are caused by read-out noise (which can dominate coherent signals in the dark areas of the field of view), as well as artifacts from JPEG compression around cosmic rays. While the despiker usually removes the cosmic rays, the compression artifacts around them can sometimes remain in the data. Most of these artifacts are removed as a result of the brightness thresholding. 

\begin{figure} 
\centerline{\includegraphics[width=0.75\textwidth,clip=true]{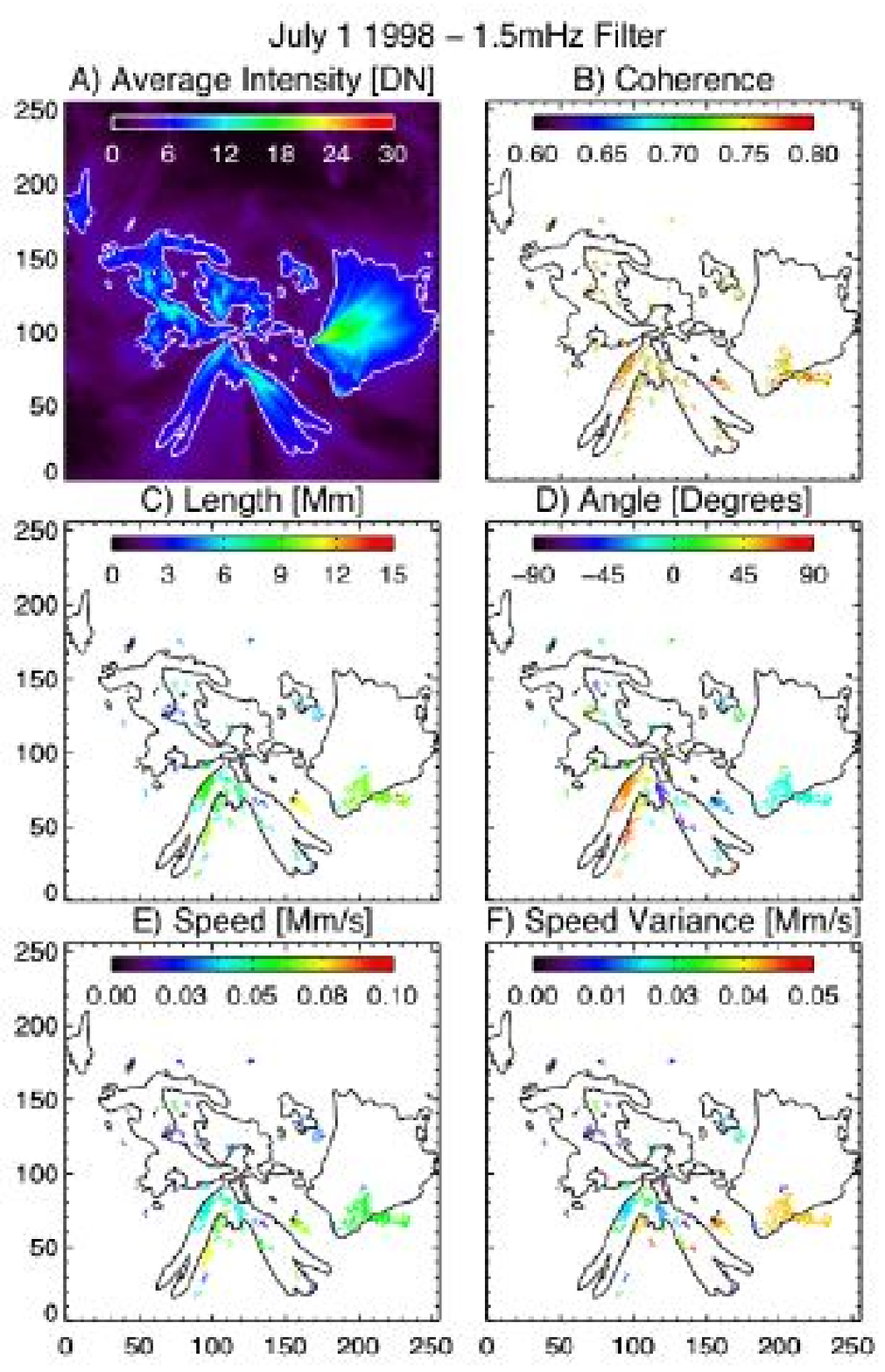}}
\caption{Final results of the coronal wave detection algorithm for the 1.5~mHz filtered timeseries for the 14 July 1998 dataset that have been ``pruned'' to reduce the appearance of false positive detections using the technique discussed in {\it Sect.}~\pref{s:rfp}. We show the average \trace{} 171\AA{} intensity image and the region averaged signal coherence, length, angle and phase speed, and the errors in the latter.}\label{fig7a}
\end{figure}

As a next step we use the map of locations and label each contiguous region in the map using the IDL routine {\texttt label\_region.pro}. We then reject all regions that contain fewer than, say, eigth pixels (superpixels of 1 $\times$ 1 arcsecond$^{2}$ in our case). This removes most of the moss regions and many of the instrumental artifacts that usually  show coherence over only a few pixels. Note that this step bases the rejection on the number of contiguous pixels in our map that have a significant number of neighbors that are highly coherent at the frequency studied. This a different from the step in {\it Sect.}~\pref{s:method} where we rejected a location from the map if it had fewer than six neighboring pixels in the island of high coherence. While these two measures are related, they are not identical. The end result of these three steps is usually a cleaner map of locations with significant oscillatory power that are coronal in nature (as opposed to instrumental). We now have a map of regions (as opposed to pixels), in which adjacent pixels have been clustered together into one region. We then calculate the average phase speed, the average error on the phase speed, the average coherence length and width, and the average angle for each of these regions. The resulting figures show maps of oscillatory ``events'' with their averaged properties (see Figure~\pref{fig7a} and~\pref{fig8a} for dataset 1 and 2 respectively). The pruning method reduces the number of ``pixels'' from 850, 409, and 129 respectively for 1.5, 3.5, and 5.5~mHz and combines them to 46, 14, and 6 regions in dataset I. Dataset II contains many more pixels with significant oscillatory power, but the pruning method reduces the number of pixels from 950, 1103, and 1150 respectively down to 91, 94, and 55 events or regions. Such a significant reduction of information in an automated fashion renders the combination of our wavetracking and pruning algorithms into an interesting option to automatically mine the \sdo/AIA data. The rejection of false positives is usually good enough so that the standard deviation of properties in each region is quite small (and is not particularly useful as a further method of rejecting artifacts).

The rejection of false positives is generally a complicated task that is quite dependent on the quality of the data (exposure times, number of cosmic ray hits, efficiency of cosmic ray removal, {\it etc.}). As a result it usually requires some fine tuning of parameters. Fortunately the quality of the \sdo/AIA data is expected to be higher and of a more constant quality than that of the TRACE data used here and so we will leave much of the ``fine tuning'' of the pruning thresholds, {\it etc.} until we have AIA data. This should greatly facilitate full and reliable automation of our data processing pipeline on the \sdo/AIA data.

\section{Discussion}\label{s:discuss}

It is clear from our results that our method reliably finds locations of significant oscillatory power in the corona. The two-step approach involves wavetracking and pruning out false-positives that can reliably and automatically detect ÒeventsÓ that are worthy of further detailed analysis. These events can be characterized by the dominant frequency, the average phase speed of the observed periodic disturbance, the average angle of propagation and the average coherence width and length (on a pixel by pixel basis), as well as the size of the cluster of coherent pixels. 

We have demonstrated that the phase speed of the propagating disturbance can be reliably used to distinguish the various wave modes that are expected to appear in the corona, with low values ($<$150~km~s$^{-1}$) of the phase speed associated with propagating slow-mode waves, and higher values ($>$150~km~s$^{-1}$) associated with either propagating fast-mode waves or standing transverse oscillations. We note however that an accurate determination of the phase speed with the available data (and its relatively slow, and non-fixed cadence) is often difficult, so that the determined phase speeds are only indicative of the real value of the phase speed. Fixed high cadence, high S/N data such as that expected from \sdo/AIA will alleviate much of this problem. In addition, we have experimented with the threshold of minimum coherence that defines the island of high coherence and find that higher thresholds generally of course reduce the size of the coherent islands, but can also lead to better determination of the phase speed. 

An additional problem that impacts the determination of the phase speed is the fact that we basically assume the island of coherence to be a roughly linear feature, such as a coronal loop. However, in certain cases, especially in fans associated with sunspots, or the large coronal dimming regions such an assumption may not be the most representative of how the propagation actually occurs. In such cases the source of the wave is relatively compact and the waves actually propagate outward in a more spherical manner, and not in a linear fashion, {\it i.e.}, along a straight line which is what our method assumes. That means that our definition of distance travelled (which we define as the distance along the long axis of the island of high coherence) does not represent a good measure of distance along the propagation direction. This can clearly lead to a less accurate determination of the phase speed, and may be why some of the measured phase speeds of the slow-mode waves in the sunspot fan are on the low side (50 km~s$^{-1}$). This problem also directly affects the determination of the other parameters, especially the propagation angle and to a lesser extent the coherence width and length. Despite these issues, the phase speeds are different enough to distinguish propagating slow-mode waves from transverse oscillations. Our algorithm routinely finds phase speeds of order several hundred km~s$^{-1}$ for the latter, which is to be expected for partially standing waves. We note that previous analyses of these transverse oscillations have indicated that there is often a significant propagating component as well as a standing component to the oscillations. This may be one of the reasons why the phase speeds we obtain are not infinite. Other reasons, of course, are related to the inherent uncertainties in determining travel times and travelled distances, as described above.

In its current form, our algorithm shows great potential for automated processing of \sdo/AIA data. It is crucial for the \sdo/AIA data that any kind of automated flagging of locations with significant oscillatory power contains so few false positives that the resulting list of events does not overwhelm the so-called ``Heliophysics Knowledge Base'' (\url{http://www.lmsal.com/helio-informatics/hpkb/}), and at the same time significantly reduces the amount of information the end-user needs to sift through before finding the subset of data that contains the oscillations of interest. As we have shown in this work, our algorithm does exactly that. It provides of order a few to a few dozen events per hour of EUV data and allows the users to immediately home in on the regions of interest that deserve more detailed study. Our algorithm takes one -- two minutes (depending on the amount of oscillations present -- we note that this does not include the time taken to derotate and coalign the image cubes) for a 256$\times$256$\times$110 datacube on a Mac Pro computer, without any effort to optimizing the FFT and correlation calculations for higher speeds. AIA data will be 2048$\times$2048 at this spatial resolution ({\it i.e.}, two $\times$ two rebinning), so that a single processor would require of order one \-- two hours to process one hours worth of data in a single passband. This means that for all coronal AIA data, a pipeline containing of order twenty CPUs would be enough to keep pace with the data flow. The removal of false positives will require some finetuning of parameters and pruning approaches ({\it e.g.}, using the cross-spectral power instead of the intensity to reject false positives) that will be investigated in the lead-up to the launch of \sdo{}, and since much will depend on the quality of the data, during the commissioning phase.

\section{Summary}\label{s:summary}

Our algorithm was applied to two datasets that contain a diverse range of oscillation or wave-like features. We were able to identify moss oscillations mostly at 1.5\--3.5~mHz, propagating slow-mode waves at 3.5\--5.5~mHz, partially standing transverse oscillations at 1.5\--5.5~mHz, as well as a large number of previously unreported apparently propagating waves at 1.5~mHz in coronal loops associated with plage (dataset I). Our algorithm has the potential to perform automated studies  of cross-wavelength correlations of oscillatory power in various EUV wavelengths \citep[171 and 195\AA{} for example, see, {\it e.g.},][]{2003AAP...404L...1K, 2001AAP...370..591R}. This will be the subject of a future paper. Our algorithm can also be run on a large number of datasets to help determine how common propagating slow-mode oscillations are in coronal loops associated with sunspots and plage. Especially the latter have been the subject of speculation since they involve the leakage of a potentially significant amount of {\it p-mode} power \citep[][]{2005ApJ...624L..61D, 2006ApJ...648L.151J} which could impact the amplitude of the {\it p-mode} spectrum \citep[][]{2007SoPh..246...53D}. 

The presence of significant oscillatory power in the 1.5~mHz passband in non-flaring active region coronal loops provides us with some interesting scientific return. Previously some oscillatory power at ten minutes had been detected in polar plumes \citep[][]{1998ApJ...501L.217D, 1999ApJ...514..441O}, but to our knowledge such power had not been reported in coronal loops in active regions. The signal is quite weak in these quiescent active-region loops, but most likely real. The fact that not all coronal loops show this propagating signal strongly suggests that slow co-alignment drifts are not the cause. We have tested this hypothesis by performing the same kind of calculations on dataset I using a variety of methods to remove the slow pointing drift that is introduced into \trace{} datasets from thermal flexing of the telescope tube \citep[][]{2000ApJ...535.1027A}. The measured propagation speeds of these 1.5 mHz waves are of order 50 km~s$^{-1}$, similar to that of the well-known 3.5 mHz propagating slow-mode oscillations. The presence of these 1.5~mHz waves is intriguing in part because their source in the lower atmosphere may well be different from that of the more well-known 3.5 mHz waves. The leakage of {\it p-modes} through the lower atmosphere has been suggested as a source for the prevalence of five minute period waves in the corona by a variety of authors \citep[][]{2005ApJ...624L..61D, 2006ApJ...648L.151J}. However, {\it p-modes} lack significant power at 1.5~mHz to drive what we see. This suggests that a different source may be working to produce these waves with ten minute periods in active region coronal loops. Recent work \citep[][]{Staus2008}, has seen the identification of horizontally propagating internal gravity waves at what is thought to be the ÒboundaryÓ between the photosphere and chromosphere. Long-standing theory \citep[{\it e.g.},][]{1967IAUS...28..429L} has these waves coupling strongly to those that propagate along magnetic field lines into the outer atmosphere. A detailed statistical study of these 1.5~mHz waves may well reveal more details about what the real source of these disturbances is and whether these acoustic gravity waves couple into the outer atmosphere. We should note that at least in this one dataset the 1.5~mHz waves are absent from the sunspot fan, and appear only in a subset of plage-associated loops. We should note that it is also possible that these 1.5~mHz waves are actually changes in the coronal loop emission itself as a result of changes in heating or cooling of the loops. Detailed visual inspection of the unfiltered and filtered movies does not allow for an unequivocal identification of the dominant process underlying the signal our algorithm finds. We intend to investigate these 1.5~mHz signals further in follow-up work by studying a wide range of different \trace{} datasets. Another new scientific finding of our numerical experiments is the presence of a significant amount of low-frequency oscillatory power in the wake of a coronal dimming. More detailed studies will be necessary to determine why these low frequencies are so prevalent in the dimming regions and whether this can be used to determine physical parameters of the dimming region itself.

\section{Conclusion}\label{s:conclude}
We have considered the problem of automatically (and robustly) isolating and extracting information about waves and oscillations observed in EUV image sequences of the solar corona with a view to near real-time application to data from the Atmospheric Imaging Array (AIA) on the {\em Solar Dynamics Observatory} (\sdo). We have found that a simple coherence / travel-time based approach detects and provides a wealth of information on transverse and longitudinal wave phenomena in the test sequences provided by the {\em Transition Region and Coronal Explorer} (\trace). The results of the search can be robustly ``pruned'' (based on diagnostic errors) to minimize false-detections such that the remainder provides reliable measurements of waves in the solar corona, with the calculated propagation speed allowing automated distinction between various wave modes and can be automatically applied to the enormous flow of data ($\approx$1Tb day$^{-1}$) that will be provided by \sdo/AIA. In addition to demonstrating the applicability of this approach we have found signatures of some interesting low frequency (1.5~mHz) oscillatory phenomena in the \trace{} test data that motivate further study.

%

\begin{figure} 
\centerline{\includegraphics[width=0.75\textwidth,clip=true]{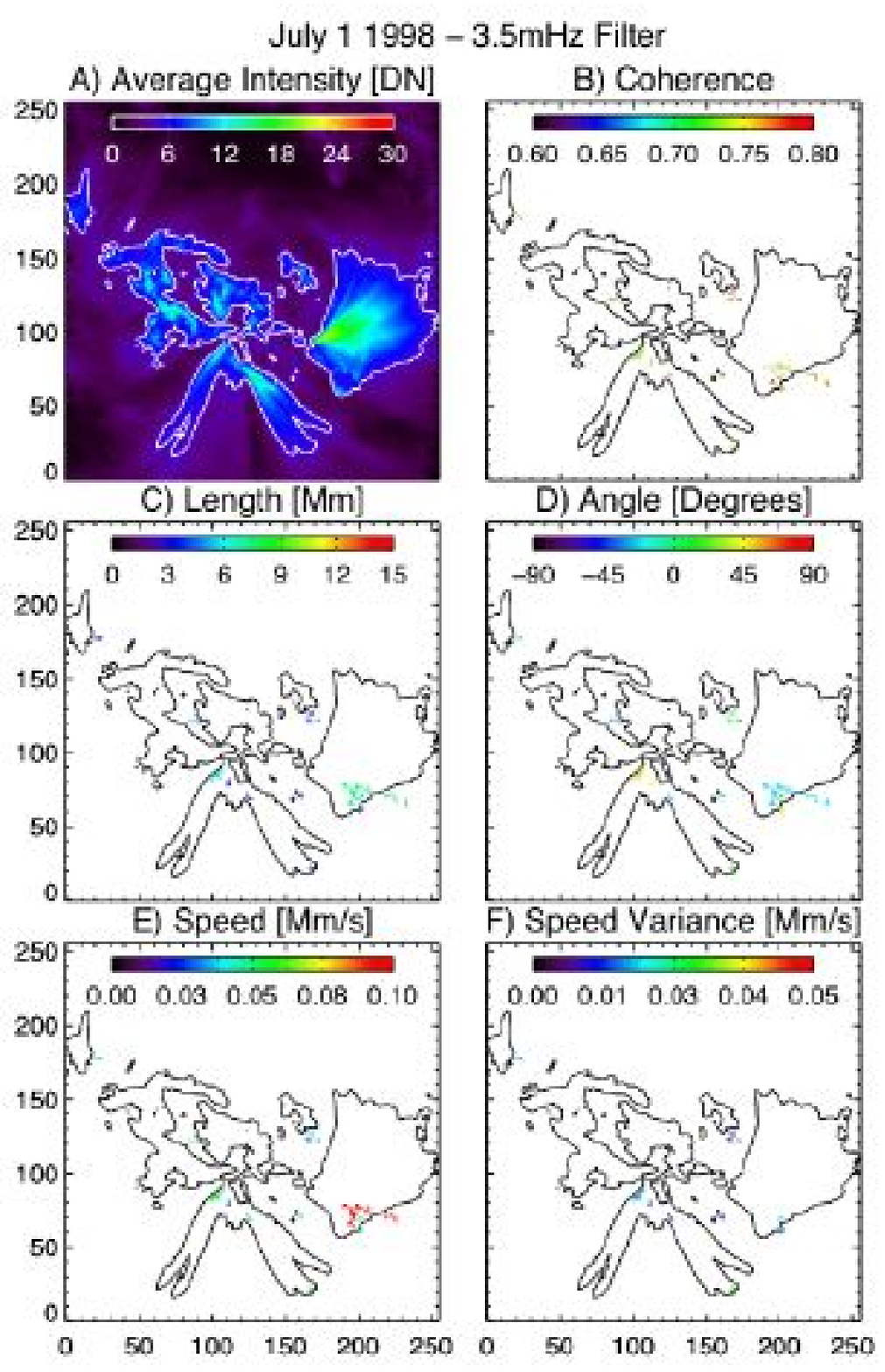}}
\caption{Final results of the coronal wave detection algorithm for the 3.5~mHz filtered timeseries for the 1 July 1998 dataset that have been ``pruned'' to reduce the appearance of false positive detections using the technique discussed in {\it Sect.}~\pref{s:rfp}. We show the average \trace{} 171\AA{} intensity image and the region averaged signal coherence, length, angle and phase speed, and the errors in the latter.}\label{fig7b}
\end{figure}

\begin{figure} 
\centerline{\includegraphics[width=0.75\textwidth,clip=true]{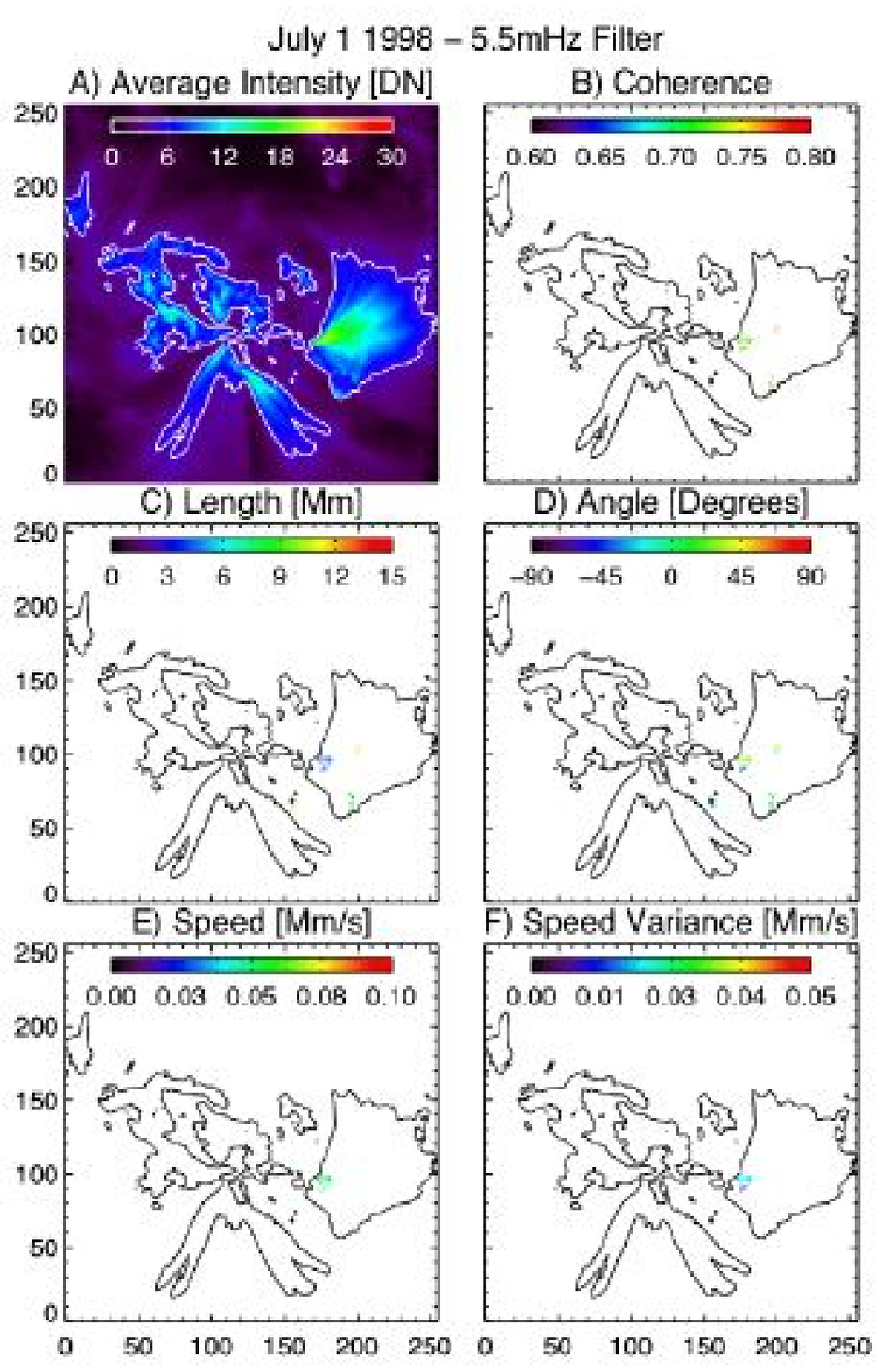}}
\caption{Final results of the coronal wave detection algorithm for the 5.5~mHz filtered timeseries for the 1 July 1998 dataset that have been ``pruned'' to reduce the appearance of false positive detections using the technique discussed in {\it Sect.}~\pref{s:rfp}. We show the average \trace{} 171\AA{} intensity image and the region averaged signal coherence, length, angle and phase speed, and the errors in the latter.}\label{fig7c}
\end{figure}

\begin{figure} 
\centerline{\includegraphics[width=0.75\textwidth,clip=true]{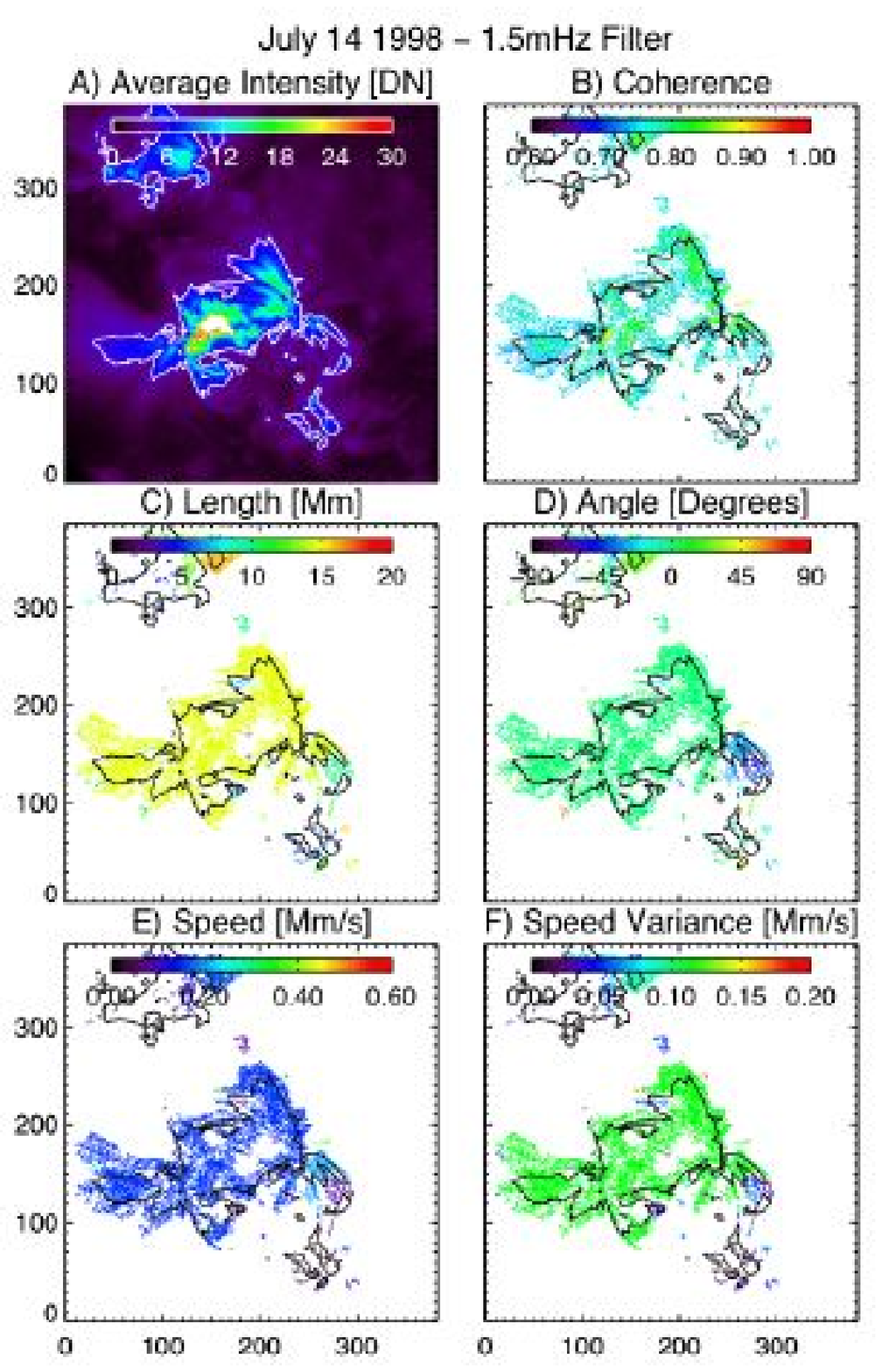}}
\caption{Final results of the coronal wave detection algorithm for the 1.5~mHz filtered timeseries for the 14 July 1998 dataset that have been ``pruned'' to reduce the appearance of false positive detections using the technique discussed in {\it Sect.}~\pref{s:rfp}. We show the average \trace{} 171\AA{} intensity image and the region averaged signal coherence, length, angle and phase speed, and the errors in the latter.}\label{fig8a}
\end{figure}

\begin{figure} 
\centerline{\includegraphics[width=0.75\textwidth,clip=true]{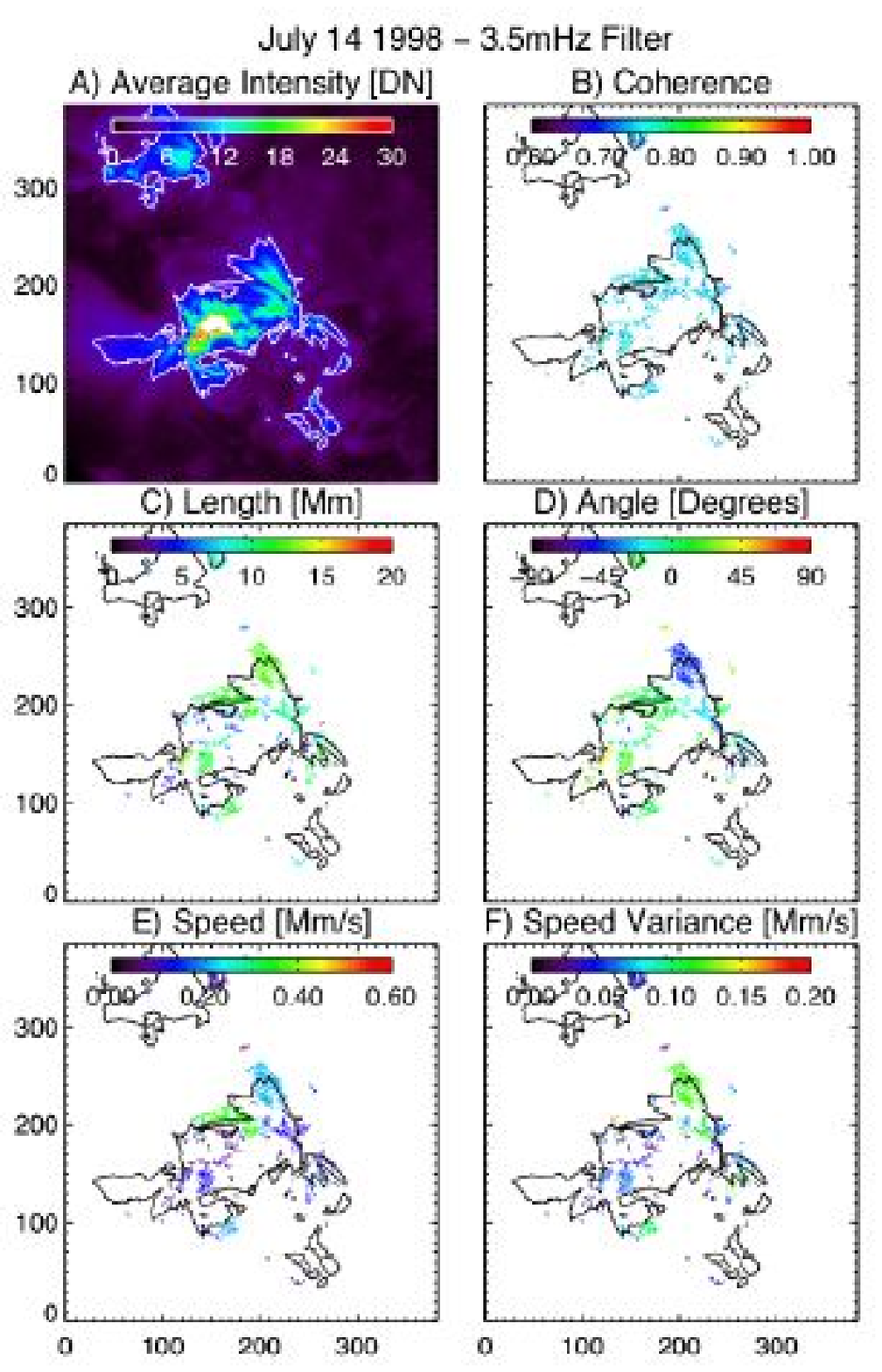}}
\caption{Final results of the coronal wave detection algorithm for the 3.5~mHz filtered timeseries for the 14 July 1998 dataset that have been ``pruned'' to reduce the appearance of false positive detections using the technique discussed in {\it Sect.}~\pref{s:rfp}. We show the average \trace{} 171\AA{} intensity image and the region averaged signal coherence, length, angle and phase speed, and the errors in the latter.}\label{fig8b}
\end{figure}

\begin{figure} 
\centerline{\includegraphics[width=0.75\textwidth,clip=true]{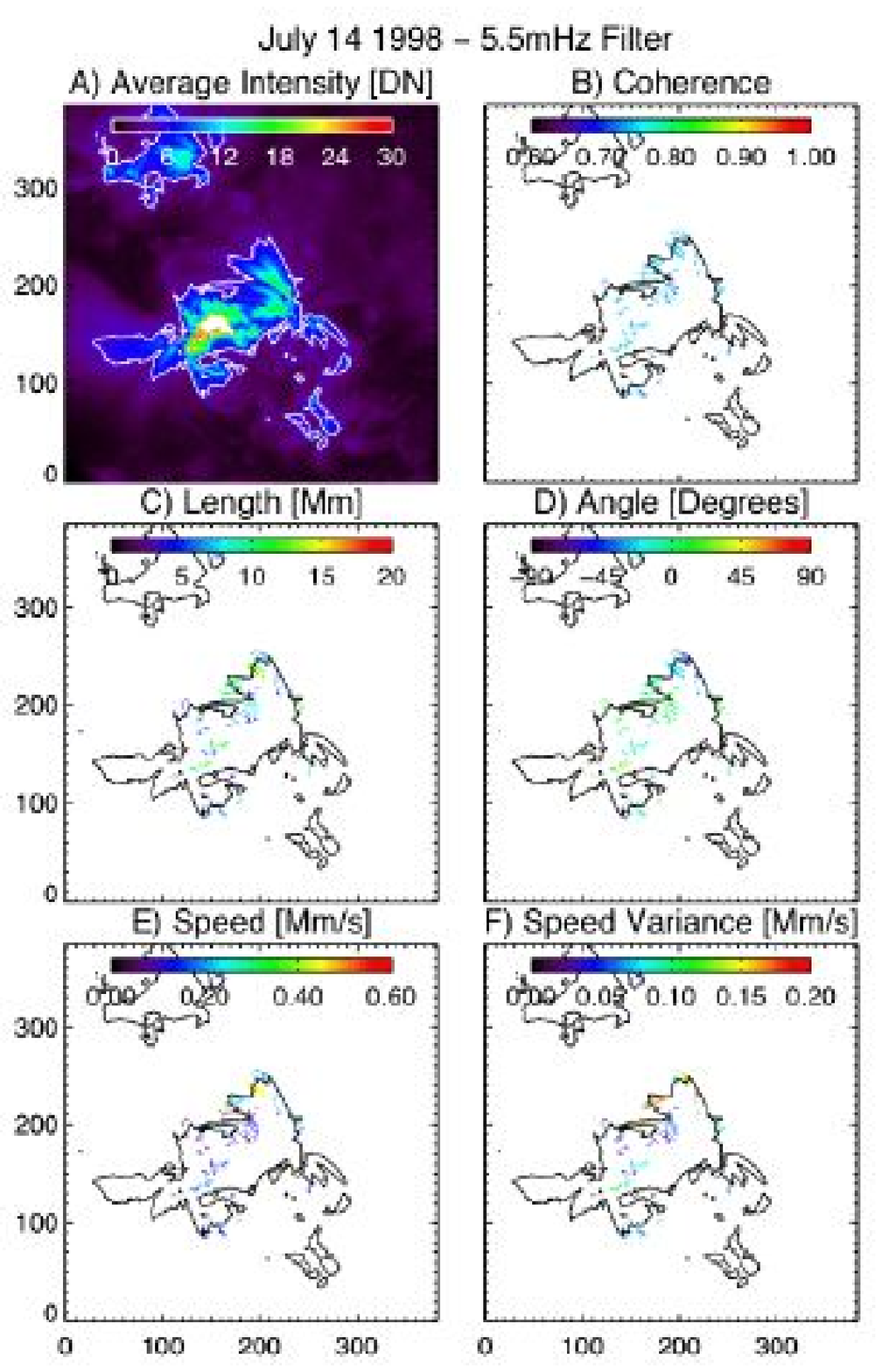}}
\caption{Final results of the coronal wave detection algorithm for the 5.5~mHz filtered timeseries for the 14 July 1998 dataset that have been ``pruned'' to reduce the appearance of false positive detections using the technique discussed in {\it Sect.}~\pref{s:rfp}. We show the average \trace{} 171\AA{} intensity image and the region averaged signal coherence, length, angle and phase speed, and the errors in the latter.}\label{fig8c}
\end{figure}

%
\acknowledgements
SWM is supported by NSF ATM-0541567, NASA NNG06GC89G; BDP by NASA grants NAS5-38099 (\trace), NNM07AA01C ({\it Hinode}) and NNG06GG79G. SWM and BDP are jointly supported by NASA grants NNX08AL22G and NNX08AH45G.

%
%
%
%
%
%

\end{document}